\documentclass[twocolumn]{article}
\usepackage{amsmath}
\usepackage{graphicx}
\usepackage{subcaption}
\usepackage{amssymb}
\usepackage{epstopdf}
\usepackage{cite}
\usepackage{authblk}
\usepackage[export]{adjustbox}
\usepackage[svgnames]{xcolor}

\newcommand{\pddf}[2]{\frac{\partial^2 #1 }{{\partial}{#2}^2}} 
\newcommand{\mbf}[1]{\mathbf{#1}} 
\newcommand{\mbb}[1]{\mathbb{#1}} 
\newcommand{\mcl}[1]{\mathcal{#1}} 
\newcommand{\trm}[1]{\textrm{#1}} 
\newcommand{\bsy}[1]{\boldsymbol{#1}} 
\newcommand{\abs}[1]{\left\lvert{#1}\right\rvert}
\newcommand{\pddxtwo}[1]{\frac{\partial^2}{\partial{#1}^2}} 
\newcommand{\argmin}{\mathop{\mathrm{argmin}}}
\usepackage{hyperref}
\hypersetup{
	colorlinks,
	linkcolor={blue},
	citecolor={blue},
	urlcolor={blue!80!black}
}
\DeclareMathOperator{\dd}{\trm{d}\!}

\newcommand{\ddx}[1]{\frac{\dd}{\dd{#1}}} 

\DeclareMathOperator{\bF}{\bsy{F}}
\DeclareMathOperator{\half}{\frac{1}{2}}
\DeclareMathOperator{\by}{\bsy{y}}
\DeclareMathOperator{\bv}{\bsy{v}}

\DeclareMathOperator{\ts}{\times}
\title{Semi-implicit Integration and Data-Driven Model Order Reduction in Structural Dynamics with Hysteresis}

\author{Bidhayak Goswami\thanks{bidhayak1728@gmail.ac.in, bidhayak@iitk.ac.in}  \and  Anindya Chatterjee\thanks{anindya100@gmail.com, anindya@iitk.ac.in}}
\date{Mechanical Engineering\\ Indian Institute of Technology Kanpur\\ \bigskip \today\\
To appear in {\em ASME Journal of Computational and Nonlinear Dynamics
}}
\begin{document}
\maketitle
\begin{abstract}
Structural damping is often empirically rate-independent {wherein the dissipative part of the stress depends on the history of deformation but not its rate of change.} Hysteresis models are popular for rate-independent dissipation; and a popular hysteresis model is the Bouc-Wen model. If such hysteretic dissipation is incorporated in a refined finite element model, then
the model involves the usual structural dynamics equations along with nonlinear nonsmooth ordinary differential equations for a large number of internal hysteretic states at Gauss points used within the virtual work calculation. For such systems, numerical integration is difficult due to both the distributed non-analytic nonlinearity of hysteresis as well as large natural frequencies in the finite element model. Here we offer two contributions. First, we present a simple semi-implicit integration approach where the structural part is handled implicitly based on the work of Pich\'e, while the hysteretic part is handled explicitly. A cantilever beam example is solved in detail using high mesh refinement. Convergence is good for lower damping and a smoother hysteresis loop. For a less smooth hysteresis loop and/or higher damping, convergence is noted to be roughly linear on average. Encouragingly, the time step needed for stability is much larger than the time period of the highest natural frequency of the structural model. Subsequently, data from several simulations conducted using the above semi-implicit method are used to construct reduced order models of the system, where the structural dynamics is projected onto a few modes and the number of hysteretic states is reduced significantly as well. Convergence studies of error against the number of retained hysteretic states
show very good results.
\end{abstract}

\section{Introduction} 
Modelling of damping in materials is a classical problem in structural dynamics, and not a fully solved one. The high dimensionality of structural finite element models combine with the non-analyticity of physically realistic damping models to produce numerical challenges in dynamic simulation. This paper makes two contributions in this area. The first contribution, which concerns simple but effective numerical integration, leads to the second contribution, which is in data-driven model order reduction.

Although the linear viscous damping model is simple and convenient, it is not always correct. For many materials subjected to periodic loading, the internal dissipation per cycle is frequency-independent \cite{kimball1927internal}. In the linear viscous damping model the dissipation per cycle is proportional to frequency. Hysteretic dissipation, which is a rate-independent \cite{muravskii2004frequency} mechanism, is preferred by many structural dynamicists because it is more realistic. However, hysteresis involves nonanalytic behavior with slope changes at every reversal of loading direction. Numerical integration for structural dynamics with hysteretic damping needs more care than with linear viscous damping. The difficulty grows greatly with finite element models wherein mesh refinement leads to high structural frequencies requiring tiny time steps; and wherein the nonanalytic hysteretic damping is finely resolved in space as well.

In this paper we first consider time integration of the vibration response for beams\footnote{%
	Our approach extends directly to frames modeled using beam elements. The approach may
	eventually be generalized to two or three dimensional elements.}
with distributed hysteretic damping, and then consider model order reduction by projecting the dynamics onto a few vibration modes. Model order reduction is not trivial in this case because the distributed hysteresis needs to be projected onto lower dimensions as well. Initial numerical solutions of the full system, using a {\em semi-implicit} integration scheme developed in this paper, are subsequently used to construct lower order models with hysteretic damping. {Our newly introduced integration scheme provides an efficient way to generate the data needed for the second part of the paper. In principle, other integration methods could be used as well, but they are either more complicated or more slow.}

\subsection{Explicit, implicit, and semi-implicit integration}\label{explicit_implicit_def}

A key idea in numerical integration of ordinary differential equations (ODEs) is outlined here for a general reader.

We consider ODE systems written in first order form, $\dot \by = \bsy{f}(\by,t)$, where $\by$ is a state vector and $t$ is time. In single-step methods, which we consider in this paper, we march forward in time using some algorithm equivalent to 
\begin{equation} \label{eqdef}
\by(t+h)=\by(t)+h \cdot \bsy{H}(t, h, \by(t), \by(t+h)).
\end{equation}
The specific form of $\bsy{H}$ above is derived from the form of $\bsy{f}(\by,t)$ and depends on the integration algorithm. The actual evaluation of $\bsy{H}$ may involve one or multiple {stages}, but that is irrelevant: the method is single-step in time. For smooth systems, $\bsy{H}$ is guided by the first few terms in Taylor expansions of various quantities about points of interest in the current time step. In such cases, as $h \rightarrow 0$, the error in the computed solution goes to zero like $h^m$ for some $m>0$. If $m>1$, the convergence is called superlinear. If $m=2$, the convergence is called quadratic. Values of $m>2$ are easily possible for smooth systems with moderate numbers of variables: see, e.g., the well known Runge-Kutta methods
\cite{chapra2011}. Unfortunately, for large structural systems, for the asymptotic $h^m$ rate to hold, $h$ may need to be impractically small.
For example, the highest natural frequency of a refined finite element (FE) model for a structure may be, say, $10^6$ Hz. This structure may be forced at, say, 10 Hz. A numerical integration algorithm that requires time steps that resolve the highest frequency in the structure, i.e., time steps much smaller than $10^{-6}$ seconds in this example, is impractical. A high order of convergence that requires time steps smaller than $10^{-6}$ is of little use. We seek accurate results with time steps much smaller than the forcing period, {but much larger than the time period of the highest natural frequency of the structure}, e.g., $10^{-2}$ or
$10^{-3}$ seconds.
To develop such practically useful algorithms, we must consider the stability of the numerical solution for larger values of $h$, i.e., $10^{-6} < h < 10^{-2}$, say.

Now consider the nature of $\bsy{H}$. If $\bsy{H}$ does not depend explicitly
on $\by(t+h)$, the algorithm is explicit. Otherwise it is implicit. For nonlinear systems, $\bsy{H}$ is a nonlinear function of
its $\by$-arguments. Then each implicit integration step requires iterative solution for $\by(t+h)$. For linear dynamics, with $\bsy{H}$ linear in
its $\by$-arguments,
the $\by(t+h)$ can be moved over to the left hand side and integration proceeds without iteration, although usually with matrix inversion. The algorithm is still called
implicit in such cases: implicitness and iteration are not the same. {An algorithm can be neither fully implicit nor fully explicit,
	and we will present one such algorithm in detail.}


\subsection{Contribution of this paper}

We present a semi-implicit approach that can be used for high dimensional finite element models with distributed hysteresis. For simplicity, we adopt the popular Bouc-Wen model {\color{magenta}\cite{bouc1967forced,wen1976method,visintin2013differential}} as our damping mechanism. After presenting and validating our numerical integration method, we present a way to obtain useful lower order models for the structure, starting from a refined FE model. A key issue is that the refined or full FE model computes hysteretic variables at a large number of Gauss points in the structure, and a smaller subset needs to be used for the model order reduction to be practical.

Thus, our contribution is twofold. First, we present a simple semi-implicit algorithm for a structural FE model with distributed hysteresis and demonstrate its convergence and utility. Second, we use this algorithm to compute some responses of the structure and use those responses to construct accurate lower order models with reduced numbers of both vibration modes and hysteretic variables. These lower order models can be used later for quick simulations under similar initial conditions or loading.

\subsection{Representative literature review}

{We begin with a review of} some numerical integration methods that are available in popular software or research papers.

Structural systems with Bouc-Wen hysteresis continue to be studied in research papers using algorithms that are not as efficient as the one we will present below. For example, as recently as 2019, Vaiana et al. \cite{vaiana2019nonlinear} considered a lumped-parameter model and used an explicit time integration method from Chang's family \cite{chang2010new}. That method requires tiny time steps: the authors used {\em one hundredth} of the time period of the highest structural mode. Such small time steps are impractical for refined FE models. Our algorithm allows much larger time steps. Thus, in the area of FE models with distributed hysteresis, our paper makes a useful contribution.

Next, we acknowledge that the commercial finite element software Abaqus \cite{abaqus2011abaqus} allows users to specify complex material responses and also to choose between explicit and implicit numerical integration options.  For many nonlinear and nonsmooth dynamic problems, explicit integration needs to be used in Abaqus. As outlined above, implicit or semi-implicit algorithms can be useful for somewhat simpler material modeling and in-house FE codes.

Considering general purpose software for numerical work, many analysts dealing with hysteresis may begin with Matlab \cite{matlab2010}. Matlab's built in function \texttt{ode15s} is designed for stiff systems but not specifically for systems with hysteresis. We have found that \texttt{ode15s} can handle ODEs from FE models with a modest number of elements and with hysteresis, but it struggles with higher numbers of elements because its adaptive time steps become too small. {In this paper we will use \texttt{ode15s} to obtain numerically accurate solutions for systems of moderate size. Results obtained from our own algorithm will then be validated against \texttt{ode15s} results. Subsequently, for larger systems, our algorithm will continue to work although \texttt{ode15s} freezes  and/or crashes.}

For those programming their own integration routines in structural dynamics, the well known Newmark method \cite{newmark1959method} from 1959 remains popular (see, e.g., \cite{bathe2007conserving,lee2017new}),
although it cannot guarantee stability and may require tiny time steps as noted in, e.g., \cite{hilber1977improved}. In that paper 
\cite{hilber1977improved} of 1977,
Hilber, Hughes and Taylor modified the Newmark method and showed unconditional stability for linear structural problems. However, their method (known as HHT-$\alpha$) can show spurious oscillations of higher modes even without hysteretic damping {(see, e.g., Fig.\ 7 of the paper by Pich\'e \cite{piche1995stable}, which we will take up below).}

An appreciation of the issues faced for full three-dimensional (3D) simulation with hysteretic damping can be gained, e.g., from the work of Triantafyllou and Koumousis \cite{triantafyllou2014hysteretic}. Their formulation is actually based on plasticity in 3D; they compare their solutions with those from Abaqus; and they include dynamics through the Newmark algorithm. Their algorithm is rather advanced for a typical analyst to implement quickly. Note that our present application is easier than theirs because we have only one-dimensional Bouc-Wen hysteresis. Additionally, our primary application here is in model order reduction. And so we develop for our current use a numerical integration approach that is simpler than that in \cite{triantafyllou2014hysteretic}.

Finally, readers interested in hybrid simulations (with an experimental structure in the loop) may see, e.g., Mosqueda and Ahmadizadeh \cite{mosqueda2007combined,mosqueda2011iterative} who used a modified version of the HHT-$\alpha$ for calculating the seismic response of a multi-degree-of-freedom structure. Purely simulation-based studies such as the present one can hopefully provide theoretical inputs into planning for such hybrid simulations in future work.

{We close this section with a brief discussion of model oder reduction. While there are very many papers on the topic, we make special note of the condensation approaches of
	Guyan \cite{guyan1965reduction} and Irons \cite{irons1965structural}, wherein finding normal modes of the full system was avoided. With growth in computational power, system eigenvectors become more easily available, and direct modal projection has become common. For a recent discussion of this and related methods, see \cite{rouleau2017comparison}. We will use straightforward modal projection for our displacement degrees of freedom. However, we will also have a large number of internal hysteretic state variables. We will present a sequential approach to selecting a subset of these hysteretic variables, using an algorithm which is related closely to a method called ``QR with column pivoting'' in \cite{golub2013matrix}. Minor differences from \cite{golub2013matrix} here are that
we work with rows and not columns, and our data matrix is not rank deficient: we separately impose a termination criterion.

Note that modal reduction, without the added complication of hysteretic damping, is well known. For example, Stringer et al.\cite{stringer2011modal,samantaray2009steady} successfully applied modal reduction to rotor systems, including ones with gyroscopic effects. Other recent examples in dynamics and vibrations can be seen in \cite{bhattacharyya2021energy,expbhattachcusumano2021}. Proper Orthogonal Decomposition (POD) \cite{chatterjee2000introduction} based reduced order models are often used in fluid mechanics: a representative sample may be found in \cite{sengupta2015enstrophy,clark2015developing,berkooz1993proper,holmes1997low}.}

\subsection{Overview of our approach}

We will begin by adopting an existing implicit scheme for linear structural dynamics without hysteresis that is both simple and significantly superior to both the Newmark and HHT-$\alpha$ methods. Our adopted scheme, due to
Pich\'e \cite{piche1995stable}, is an L-stable method from the Rosenbrock family (see \cite{wanner1996solving} and references therein) which is stable, implicit without iteration for linear structures, and second order accurate. For smooth nonlinear systems of the form
$$\mbf{M} \ddot{\bsy{x}} + \bsy{f}(\bsy{x}, \dot{\bsy{x}}) = \bsy{0},$$
Pich\'e  suggests a one-time linearization of $\bsy{f}$ at the start of each time step. We will not use that linearization step because hysteresis is nonsmooth.

Here we extend Pich\'e's formulation to include hysteretic damping in the
otherwise-linear structural dynamics.
In our method the stiffness and inertia terms are integrated implicitly (without iteration, being linear) and the nonsmooth hysteresis is monitored at Gauss points \cite{maiti2018vibrations} and treated with explicit time-integration. Thus, overall, our method is semi-implicit. Our proposed approach, when applied to a refined FE model of an Euler-Bernoulli beam with distributed Bouc-Wen hysteretic damping, is easy to implement and can be dramatically faster than Matlab's \texttt{ode15s}. In fact, it continues to work well beyond refinement levels where \texttt{ode15s} stops working.

We emphasize that \texttt{ode15s} is not really designed for such systems. We use it here because it has adaptive step sizing and error estimation: when it does work, it can be highly accurate. For this reason, to examine the performance of our algorithm for modest numbers of elements, we will use results obtained from \texttt{ode15s} with a tight error tolerance.
With higher number of elements, \texttt{ode15s} fails to run because the time steps required become too small. 

Having shown the utility of our proposed semi-implicit integration algorithm, we will turn to our second contribution of this paper, namely model order reduction.
Such Reduced Order Models (ROMs) can save much computational effort. In the physical system, if only a few lower modes are excited, the displacement vector can be approximated as a time-varying linear combination of those modes only. A remaining issue here is that the number of Gauss points used for the hysteresis can be reduced too, and that is where we offer an additional contribution.

In recent work related to ours, for an Euler-Bernoulli beam with hysteretic dissipation, Maity et al. \cite{maiti2018vibrations} used the first few undamped analytically obtained modes to approximate the solution and performed the virtual work integration using a few Gauss points chosen over the full domain. However, they used a different hysteresis model motivated by distributed microcracks. Furthermore,
in our finite element model, virtual work integrations are performed over individual elements and not the whole domain. The number of Gauss points in the finite element model increases with mesh refinement. When we project the response onto a few modes, we can explicitly reduce the number of Gauss points retained as well, and a practical method of doing so is one of the contributions of this paper.

In what follows, we present the finite element model in section \ref{fe_formulation}, outline the numerical integration algorithm in section \ref{SE_scheme}, develop the approach for model order reduction in section \ref{ROM_sec}, and present our results in section \ref{results}.

\section{Governing equations and Finite Element formulation}\label{fe_formulation}
\begin{figure}[ht!]
	\centering 
	\includegraphics[scale=0.5,center]{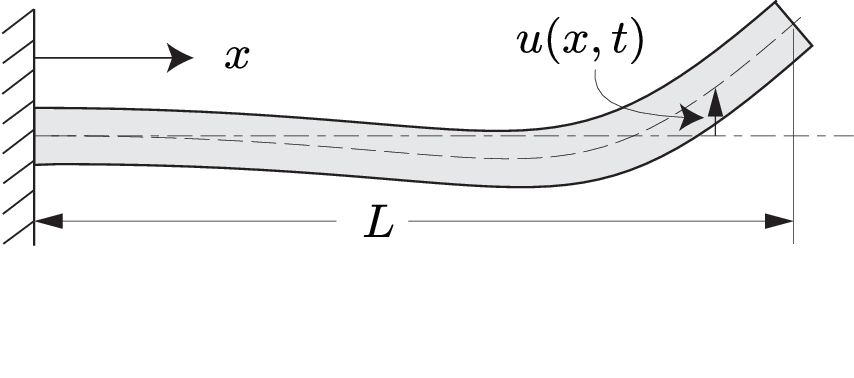}
	\caption{A cantilever beam.}
		\label{beam_fig1}
\end{figure}

The governing equation of the deflection $ {u}({x},{t}) $ of a cantilever beam (shown in Fig.\ (\ref{beam_fig1})) with a dissipative bending moment $ M_{\trm{d}}=\gamma_{\trm{h}} z(x,t) $  is
\begin{equation}\label{eq_gov}
			\rho A \pddf{{u}}{{t}}+\pddxtwo{{x}}\left(E I \pddf{
			{u}}{{x}}+{\gamma}_{\trm{h}} z\right)=0, 
\end{equation}
where
the beam's material density, cross section and flexural rigidity are
$\rho$, $A$ and $EI$ respectively. The parameter $ \gamma_{\trm{h}} $ is the strength of hysteretic dissipation. 
The hysteretic damping variable $ z $ is defined at each $ x $-location along the beam length, is governed pointwise by the Bouc-Wen model, and is driven pointwise by the local curvature
$$ {\chi}({x},{t})\approx \pddf{{u}}{{x}}$$
in the governing equation
	\begin{equation}\label{Bouc_wen}
	    \dot{z}=\left({\bar{A}}-{\alpha}\: \trm{sign}\left(\dot{\chi} \,z\right) \abs{z}^{n_{\trm{h}}}-{\beta} \abs{z}^{n_{\trm{h}}}\right)\dot{\chi}
	\end{equation}
	where the Bouc-Wen model parameters satisfy    
\begin{equation}
\label{BWp}
	\bar{A}>0,\,\alpha>0,\,\beta\in (-\alpha,\alpha)\:\trm{and}\:n_{\trm{h}}>0.
\end{equation}
The parameters in Eq.\ (\ref{BWp}) and the hysteretic variable  $z$ are dimensionless. 
The dissipation strength parameter $\gamma_{\trm{h}}$ has units of Nm (i.e., units of moments).

The FE model involves beam elements for displacement variables, virtual work calculations for the hysteretic moment through domain integrals based on Gauss points within each element, and ODE solution in time. We use the Galerkin method wherein we look for an admissible solution $\hat{u}({x},t)$ so that the residual,
	\begin{equation}
	    \mcl{R}(\hat{u})=\rho A \pddf{\hat{u}}{t}+\pddxtwo{x}\left(E I \pddf{\hat{u}}{x}+\gamma_{\trm{h}} z\right),
	\end{equation}
	is orthogonal to basis functions for the space of $\hat{u}$. Mathematically,
	\begin{equation}\label{galerkin}
		<\mcl{R}(\hat{u}),\phi_i> \, =\int_{\Omega}\mcl{R}(\hat{u})\phi_i({x}) \dd \Omega=0,
	\end{equation}
	where
$$ <f_1,f_2> \, = \int_{\Omega} f_1({x}) f_2({x}) \dd \Omega$$  {is the inner product between two functions, \(\Omega\) is the spatial domain over which the basis functions are defined,}
and $\phi_i$ is the $i^{\rm th}$ basis function.

\subsection{Element matrices and virtual work integrals}

The elemental stiffness and inertia matrices are routine and given in many textbooks: each beam element has two nodes, and each node has one translation and one rotation. The hysteresis variable $z$ is a scalar quantity distributed in space. However, it enters the discretized equations through integrals, and so $z$ values only need to be known at Gauss points within each element. The evolution of $z$ at the Gauss points is computed using Eq.\ (\ref{Bouc_wen}),
and is thus driven by displacement variables. 
Some further details are given in appendix \ref{appB}.

\subsection{Global equations}

After assembly we obtain a system of equations of the form
\begin{subequations}
	\begin{equation}\label{full_model}
			\underset{\trm{Mass}}{\mbf{M}\,\ddot{\bsy{q}}}+\underset{\trm{Stiffness}}{\mbf{K}\,\bsy{q}}+\underset{\trm{Hysteresis}}{\mbf{A}\,\bsy{z}}=\underset{\trm{Forcing}}{\bsy{f}_0(t)},
	\end{equation}
where \(\mbf{A}\in \mbb{R}^{N\ts n_{\trm{g}} n_{\trm{e}}}\), \(\bsy{q}\in \mbb{R}^N\), and \(\bsy{z}\in\mbb{R}^{n_{\trm{g}} n_{\trm{e}}}\). {The hysteresis rate equations are given by}
	\begin{equation}\label{full_model_zdot}
			\dot{\bsy{z}}=(\bar{A}-\alpha\, \trm{sign}(\dot{\bsy{\chi}}\circ\bsy{z})\circ\abs{\bsy{z}}^{\circ\, n_{\trm{h}}}-\beta \abs{\bsy{z}}^{\circ\, n_{\trm{h}}})\circ\dot{\bsy{\chi}},\,\, \bsy{\chi}=\mbf{B} \bsy{q},
	\end{equation}
\end{subequations}
where \(\mbf{B}\in \mbb{R}^{n_{\trm{e}} n_{\trm{g}}\ts N}\) and the different symbols mean the following: 
\begin{enumerate}
\item $\bsy{q}$ is a column vector of nodal displacements and rotations for $n_{\trm{e}}$ beam elements, with $N = 2 n_{\trm{e}}$ for a cantilever beam,
\item $\mbf{M}$ and $\mbf{K}$ are mass and stiffness matrices of size $2\,n_{\trm{e}} \times 2\,n_{\trm{e}}$,
\item $\bsy{z}$ is a column vector of length $n_{\trm{g}} n_{\trm{e}}$ from $n_{\trm{g}}$ Gauss points per element,
\item $(\cdot \circ \cdot)$ and $(\cdot)^{\circ\, (\cdot)}$ denote elementwise multiplication and exponentiation,
\item  $\mbf{A}$ is a matrix of weights used to convert $\bsy{z}$ values into virtual work integrals,
\item  $\bsy{\chi}$ is a column vector of curvatures at the Gauss points, 
\item  $\mbf{B}$ maps nodal displacements and rotations $\bsy{q}$ to curvatures $\bsy{\chi}$ at the Gauss points, and
\item  $\bsy{f}_0(t)$ incorporates applied forces.
\end{enumerate}
In Eq.\ (\ref{full_model}) above we can include viscous damping by adding $\mbf{C}\dot{\bsy{q}}$, for some suitable $\mbf{C}$, on the left hand side. {As this paper focuses on hysteretic damping, we have taken \({\bf C}={\bf 0}.\)}
	
\section{Time integration}
\label{SE_scheme}

We will develop a semi-implicit method adapted from Pich\'e's \cite{piche1995stable} work. Equation (\ref{full_model}) has a structural part (stiffness and inertia) and a hysteresis part. The structural part is integrated implicitly (section \ref{implicit_integration}) and the hysteresis part marches in time, following an explicit algorithm which takes care of the nonsmooth slope changes due to zero-crossing of the time derivative of the curvature (section \ref{explicit_integration}).
In section \ref{ROM_sec}, our semi-implicit algorithm will be used to generate a large amount of data which will be used to construct reduced order models.

\subsection{Implicit integration for the structural part}
\label{implicit_integration}

Pich\'e's algorithm uses a numerical parameter $1-\frac{1}{\sqrt{2}}$ which is written as $\gamma$ for compact presentation. The symbol should not be confused for Euler's constant.
We now proceed as follows. This subsection presents implicit integration for the structural part alone: the hysteretic variable vector $\bsy{z}$ is not integrated in an implicit step: this compromise simplifies the algorithm greatly.

\begin{enumerate}
	    \item Define
	\[\bsy{z}(t_0)=\bsy{z}_0,\quad \bsy{F}_0=-\mbf{A}\bsy{z}_0+\bsy{f}_0(t_0),\]\[ \by_0=\bsy{q}(t_0),\quad \bv_0=\dot{\bsy{q}}(t_0) ,\quad \dot{\bsy{\chi}}_0=\mbf{B}\bv_0,\]
	\[\dot{\bsy{z}}_0= \left(\bar{A}-\alpha\,\trm{sign}(\dot{\bsy{\chi}}_0\circ\bsy{z}_0)\circ \abs{\bsy{z}_0}^{\circ n_{\trm{h}}}-\beta \abs{\bsy{z}_0}^{\circ n_{\trm{h}}}    \right)\circ \dot{\bsy{\chi}}_0, \]\[ \dot{\bsy{F}}_0=-\mbf{A}\dot{\bsy{z}}_0+\ddx{t}\bsy{f}_0(t)\bigg\rvert_{t=t_0},\]
	\[\bsy{r}_0=\mbf{K}\by_0 + \mbf{C}\bv_0  \]
	(The \(\mbf{C}\bv_0\) is dropped if linear viscous damping is no included.)
	\item Define
	\[\tilde{\mbf{M}}=\mbf{M}+\gamma h\mbf{C}+(\gamma h)^2\,\mbf{K}. \]
	\item Define (first stage)
	\[\tilde{\bsy{e}}=h\,{\tilde{\mbf{M}}}^{-1}\left(\bsy{F}_0-\bsy{r}_0+h\gamma(\dot{\bsy{F}}_0-\mbf{K}\bsy{v}_0) \right).\]
	\[\tilde{\bsy{d}}=h(\bsy{v}_0+\gamma\tilde{\bsy{e}}).\]
	\item Define (second stage)
	\[
	\bF_{\half}=-\mbf{A}\left(\bsy{z}_0+\frac{h}{2}\dot{\bsy{z}}_0\right)+\bsy{f}_0\left(t_0+\half h\right), \]\[ \bsy{r}_{\half}=\mbf{K}\left(\by_0+\half\tilde{\bsy{d}}\right)+\mbf{C}\left(\bv_0+\half\tilde{\bsy{e}} \right),\]
	\[
	\bsy{e}=h\tilde{\mbf{M}}^{-1}\left(\bF_{\half}-\bsy{r}_{\half}+(h\gamma)\left(2\gamma-\half\right)\mbf{K}\tilde{\bsy{e}}+\gamma\mbf{C}\tilde{\bsy{e}}\right),
	\]
	and
	\[
	\bsy{d}=h\left(\bsy{v}_0+\left(\half-\gamma\right)\tilde{\bsy{e}}+\gamma\bsy{e}
		\right).
	\]
	\item Finally 
	\[
	\bsy{y}(t_0+h)=\by_0+\bsy{d},\qquad\bv(t_0+h)=\bv_0+\bsy{e}.
	\]
	\item Define
	\[
	\dot{\bsy{\chi}}_1=\dot{\bsy{\chi}}(t_0+h)=\mbf{B}(\bv_0+\bsy{e})
	\]
\end{enumerate}
Note that the assignment of $ \dot{\bsy{\chi}}_1 $ to $ \dot{\bsy{\chi}}(t_0+h) $ above is tentative at this stage; if there is a sign change, we will make a correction, as discussed in the next subsection.

\subsection{Explicit integration for the hysteretic part}
\label{explicit_integration}

Due to the {nonanalyticity} of hysteresis models, we integrate the hysteresis part using an explicit step. There are slope changes in the hysteretic response whenever $\dot{\chi}$ at any Gauss point crosses zero in time. We will accommodate the sign change of $\dot{\chi}$ within an explicit time step in a second sub-step, after first taking a time step assuming that there is no sign change.
Since each $\dot{z}_i$ depends on $ \dot{\chi}_i $ only, accounting for zero crossings can be done individually for individual Gauss points and after such a preliminary step.

\begin{figure}[h!]
	\centering
	\includegraphics[scale=0.61]{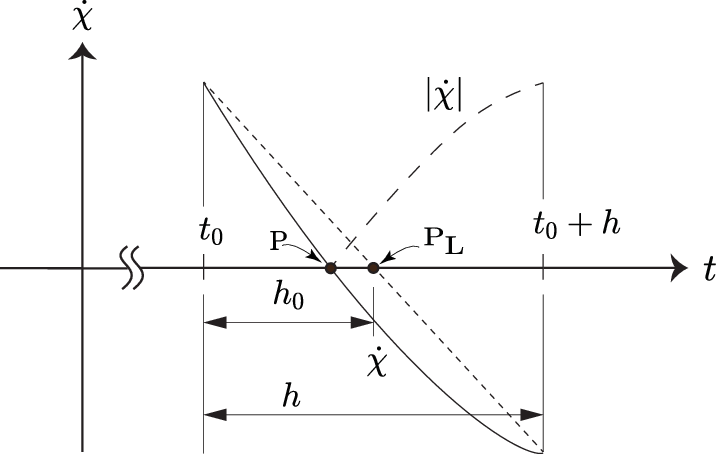}
	\caption{Zero-crossing of $ \dot{\chi} $ at any particular Gauss point within a time step. For details on associated quantities shown, see the main text.}
	\label{crossing}
\end{figure}	
	
We first check elementwise and identify the entries of $ \dot{\bsy{\chi}}_0 $ and $ \dot{\bsy{\chi}}_1 $ that have same sign, and construct sub-vectors $ \dot{\bsy{\chi}}_{0\trm{u}} $ and $ \dot{\bsy{\chi}}_{1\trm{u}} $ out of those entries. The corresponding elements from the vector $ \bsy{z}_0 $ are used to construct a sub-vector $ \bsy{z}_{0\trm{u}} $. Here the subscript ``$ {\rm u} $" stands for ``unchanged''. These elements can be used in a simple predictor-corrector step for improved accuracy as follows.
	    \[
	    \bsy{s}_1=\left(\bar{A}-\alpha \,\trm{sign}(\dot{\bsy{\chi}}_{0\trm{u}}\circ\bsy{z}_{0\trm{u}})\circ\abs{\bsy{z}_{0\trm{u}}}^{\circ n_{\trm{h}}}-\beta \abs{\bsy{z}_{0\trm{u}}}^{\circ n_{\trm{h}}} \right)\circ \dot{\bsy{\chi}}_{0\trm{u}}, \]
\[ \bsy{z}_{\trm{iu}}=\bsy{z}_{0\trm{u}}+h \bsy{s}_1,
\]
	    \[
	    \bsy{s}_2=\left(\bar{A}-\alpha \,\trm{sign}(\dot{\bsy{\chi}}_{1\trm{u}}\circ\bsy{z}_{\trm{iu}})\circ\abs{\bsy{z}_{\trm{iu}}}^{\circ n_{\trm{h}}}-\beta \abs{\bsy{z}_{\trm{iu}}}^{\circ n_{\trm{h}}} \right)\circ \dot{\bsy{\chi}}_{1\trm{u}},\]
and
\begin{equation}
\label{eq:hst}
	    \bsy{z}_{1\trm{u}}=\bsy{z}_{0\trm{u}}+h\, \frac{(\bsy{s}_1+\bsy{s}_2)}{2}.
\end{equation}

Next, we address the remaining entries of $\dot{\bsy{\chi}}_0$ and $\dot{\bsy{\chi}}_1$, those that have crossed zero and flipped sign.
We  construct sub-vectors $\dot{\bsy{\chi}}_{0\trm{f}}$ and $\dot{\bsy{\chi}}_{1\trm{f}}$ out of them, where the ``${\rm f}$'' subscript stands for ``flipped''. The corresponding elements from the vector $\bsy{z}_0$ are used to construct sub-vector
 $\bsy{z}_{0\trm{f}}$. Using linear interpolation to approximate the zero-crossing instants within the time step, we use
	    \[
	    \bsy{h}_0=-h\,\frac{\dot{\bsy{\chi}}_{0\trm{f}}}{(\dot{\bsy{\chi}}_{1\trm{f}}-\dot{\bsy{\chi}}_{0\trm{f}})}\quad \trm{(defined
elementwise)}.
	    \]
	    We define
	    $$
	    \bsy{s}_1=\left(\bar{A}-\alpha \,\trm{sign}(\dot{\bsy{\chi}}_{0\trm{f}}\circ\bsy{z}_{0\trm{f}})\circ\abs{\bsy{z}_{0\trm{f}}}^{\circ n_{\trm{h}}}-\beta \abs{\bsy{z}_{0\trm{f}}}^{\circ n_{\trm{h}}} \right)\circ \dot{\bsy{\chi}}_{0\trm{f}},$$
and
$$ \bsy{z}_{\trm{mf}}=\bsy{z}_{0\trm{f}}+\bsy{h}_0 \circ \frac{\bsy{s}_1}{2},$$
which is consistent with Eq.\ (\ref{eq:hst}) because $\dot{\bsy{\chi}} = \bsy{0}$ at the end of the sub-step.
Next we complete the step using
	    \[
	    \bsy{s}_2=\left(\bar{A}-\alpha \,\trm{sign}(\dot{\bsy{\chi}}_{1\trm{f}}\circ\bsy{z}_{\trm{mf}})\circ\abs{\bsy{z}_{\trm{mf}}}^{\circ n_{\trm{h}}}-\beta \abs{\bsy{z}_{\trm{mf}}}^{\circ n_{\trm{h}}} \right)\circ \dot{\bsy{\chi}}_{1\trm{f}}
	    \]
	   and
	    \[
	    \bsy{z}_{1\trm{f}}=\bsy{z}_{\trm{mf}}+(\bsy{h}-\bsy{h}_0)\circ \frac{\bsy{s}_2}{2}
	    \]
	    where $\bsy{h}$ is a vector of the same dimensions as $\bsy{h}_0$ and with all elements equal to $h$.
Finally, having the incremented values $\bsy{z}_1$ at all locations, those with signs unchanged and those with signs flipped, we use
	    \[ \bsy{z}(t_0+h)=\bsy{z}_1.  \] 
We clarify that the explicit integration of the hysteretic variables, as outlined in this subsection, is a compromise adopted to avoid difficulties due to the nonanalyticity of the hysteresis model. Continuing to integrate the inertia and stiffness parts explicitly will still allow us to use usefully large steps, as will be seen in section \ref{results}.

Having the numerical integration algorithm in place, however, we must now turn to the second contribution of this paper, namely model order reduction.

	\section{Model order reduction}\label{ROM_sec}
	
For reduced order modeling, we must reduce both the number of active vibration modes as well as the number of Gauss points used to compute the hysteretic dissipation. Of these two, the first is routine.

	\subsection{Reduction in the number of vibration modes}\label{modal_rom}
	
	The undamped normal modes and natural frequencies are given by the eigenvalue problem
	\begin{equation}
		(\mbf{K}-\omega^2\mbf{M})\bsy{v}=\bsy{0}
	\end{equation}
	for $N$ eigenvector-eigenvalue pairs $(\bsy{v}_i,\omega_i)$. In a finely meshed FE model, $N$ is large. In such cases, we may compute only the first several such pairs using standard built-in iterative routines in software packages.

The $N$ dimensional displacement vector $\bsy{q}$ is now approximated as a linear combination of  $r \ll N $ modes. To this end, we construct a
matrix $\mbf{R}$ with the first $r$ eigenvectors as columns so that
	\begin{equation}\label{modal_approx}
		\bsy{q}(t)\approx\sum_{k=1}^{r}\bsy{v}_k {\xi}_k(t)=\mbf{R}\bsy{\xi}(t), 
	\end{equation}
where $\mbf{R}=[\bv_1\,\;\bv_2\,\dots\, \bv_r]$, and where the elements of $\bsy{\xi}(t)$ are called modal coordinates.
	Substituting Eq.\ (\ref{modal_approx}) in Eq.\ (\ref{full_model}) and pre-multiplying with the transposed matrix $\mbf{R}^{\top}$, we obtain
	\begin{equation}\label{modal_rom_1}
		\mbf{R}^{\top}\mbf{M}\mbf{R}\:\ddot{\bsy{\xi}}+	\mbf{R}^{\top}\mbf{K}\mbf{R}\:\bsy{\xi}+\mbf{R}^{\top}\mbf{A}\bsy{z}=\mbf{R}^\top \bsy{f}_0(t),  
	\end{equation}
obtaining equations of the form
	\begin{subequations}
		\begin{equation}\label{modal_rom_2}
			\tilde{\mbf{M}}\:\ddot{\bsy{\xi}}+\tilde{\mbf{K}}\:\bsy{\xi}+\mbf{R}^{\top}\mbf{A}\bsy{z}=\mbf{R}^\top\bsy{f}_0(t),
		\end{equation}
		\begin{equation}\label{modal_rom_z}
			\dot{\bsy{z}}=(\bar{A}-\alpha\: \trm{sign}(\dot{\bsy{\chi}}_{\trm{a}}\circ\bsy{z})\circ\abs{\bsy{z}}^{\circ\, n_{\trm{h}}}-\beta \abs{\bsy{z}}^{\circ\, n_{\trm{h}}})\circ\dot{\bsy{\chi}}_{\trm{a}},\,\, \dot{\bsy{\chi}}_{\trm{a}}=\mbf{B}\mbf{R}\dot{\bsy{\xi}}.
		\end{equation}
	\end{subequations}
In the above, due to orthogonality of the normal modes, the matrices $\tilde{\mbf{M}}$ and $\tilde{\mbf{K}}$ are diagonal.
Also, $\bsy{\chi}_{\trm{a}}$ is an approximation of the original $\bsy{\chi}$ because we have projected onto a few vibration modes, but it still has the same number of elements as $\bsy{\chi}$ and requires numerical integration of the same number of nonsmooth hysteresis equations. A reduction in the number of hysteretic variables is necessary.

\subsection{Reduction in the number of hysteretic variables}\label{hyst_var_reduction_section}\label{choose_z_subset}
The system still has a large number ($n_{\trm{g}} n_{\trm{e}}$) of hysteretic damping variables. These are arranged in the elements of  $\bsy{z}$ in Eq.\ (\ref{full_model}). Selecting a smaller set of basis vectors and projecting the dynamics of the evolving $\bsy{z}$ onto those has analytical difficulties.
Here we adopt a data-driven approach to select a submatrix of $\bsy{z}$, say
\begin{equation}\label{z_s}
	\bsy{z}_{\trm{s}}=[z_{j_1}\, z_{j_2}\,\dots,z_{j_m}]^{\top},\, m\ll n_{\trm{g}} n_{\trm{e}}.
\end{equation}
The indices $j_1,\,j_2,\,\dots j_m $ must now be chosen form the set \(\{1,2,\dots,n_{\rm e}n_{\rm g}\}\), along with a matrix ${\bf P}$, such that
\begin{equation}\label{P_condition}
	\mbf{P}\bsy{z}_{\trm{s}} (t) \approx \mbf{R}^{\top}\mbf{A}\bsy{z}(t).
\end{equation}
Working with that reduced set of hysteretic variables, we will be able use a reduced set of driving local curvatures
\begin{equation}\label{chi_s}
	\dot{\bsy{\chi}}_{\trm{s}}=[\dot{\chi}_{\trm{a}_{j_1}}\,\dot{\chi}_{\trm{a}_{j_2}}\,\dots\,\dot{\chi}_{\trm{a}_{j_m}}]^{\top},
\end{equation}
a submatrix of $\dot{\bsy{\chi}}_{\trm{a}}$. In other words, we will work with a subset of the original large number of Gauss points used to
compute virtual work integrals for the hysteretic dissipation.
In our proposed data-driven approach, selection of the indices $j_1,\,j_2,\,\dots j_m$ is the only significant issue. The matrix ${\bf P}$ can then
be found by a simple matrix calculation.

However, for implementing our approach, we must first generate a sufficiently large amount of data using numerical integration of the full equations with representative initial conditions and/or forcing. 

{These initial conditions are randomly generated as follows. The hysteretic states are uniformly distributed random numbers in the interval (0, 0.1). For the displacement and rotation degrees of freedom, the initial conditions have nonzero values along only the first three modes. These initial conditions, in all cases, are scaled to make the tip displacement to be 2 cm. The three modal coordinate values are three randomly chosen, positive, {independent and identically distributed} numbers, subsequently rearranged so that the first mode had the largest displacement and the third mode had the smallest displacement. The initial values of the velocity degrees of freedom are taken to be zero. In applications, users who have other preferred criteria for choosing representative initial conditions can freely use them with no consequences for the rest of the algorithm.

Having selected initial conditions, we must simulate the system by integrating the equations forward in time.} To that end, we use the semi-implicit integration method presented in section (\ref{SE_scheme}). 

The data generation process is relatively slow, but once it is done and the reduced order model is constructed, we can use it for many subsequent quicker calculations.
Beyond its academic interest, our approach is practically useful in situations where the reduced order model will be used repeatedly after it is developed.

{It is perhaps not strictly necessary to place both of our contributions within one paper, but in our implementation the second part relies heavily on the first part; moreover, the second part is a relatively small and simple application of an efficient method to generate the needed data.}
	
\subsubsection{Selecting a subset of hysteretic variables}	
\label{z_selection}

For data generation, we numerically integrate the full system, over some time interval $[0,T]$ with $N_t $ points in time. This process is carried out $N_{\trm{s}}$ times with different random representative initial conditions as described above. The integration duration $T$ is chosen as 1 second,
which gives several cycles of oscillation of the lowest mode along with some significant
decay in oscillation amplitude.
For each such solution with a random initial condition, upon arranging the $z$ variables' solutions side by side in columns, a matrix of size $n_{\trm{g}} n_{\trm{e}}\ts N_t$ is generated. This matrix is divided by its Frobenius norm, i.e., the square root of the sum of squares of all elements. We stack $N_{\trm{s}}$ such normalized matrices, one from each simulation, side by side to form a bigger matrix $\mbf{Z}$ of size ${n_{\trm{g}} n_{\trm{e}} \ts N_t N_{\trm{s}}}$. Clearly, the dimension of the row-space of $\mbf{Z}$ is at most $n_{\trm{g}}n_{\trm{e}}$, because that is the total number of rows. Identification of a useful
subset of hysteretic variables is {equivalent to identifying a useful subset of the rows of \({\bf Z}\). Therefore the problem we face now is that of} selecting rows of $\mbf{Z}$ which contain a high level of independent information.
	
Selecting a finite subset of the rows of $\mbf{Z}$ is a combinatorial optimization problem. For example, if we start with $n_{\trm{g}} n_{\trm{e}} = 900$ and want to select a subset of size, say, 100, then the total number of subsets possible is
{extremely large and only a modest number of them can realistically be checked. Sometimes low-rank approximations to data matrices can be obtained using} the proper orthogonal decomposition (POD) \cite{chatterjee2000introduction}, {but that technique is not useful here because it does not select a subset of rows}.

Here, for a simple practical solution, we use a greedy algorithm {that is closely related to a use of the QR decomposition (rank deficient least squares solutions) discussed in \cite{golub2013matrix}. Although it is not guaranteed to give the best solution, we will see that it does give a reasonably good solution.}

\begin{enumerate}
\item Of all the rows of $\mbf{Z}$, we first select the one with the largest 2-norm and set it aside; we also record its index (or row number), which we call
$j_1$. {We scale the \(j_1^{\rm th}\) row to unit norm, calculate its inner products with the rest of the rows of \({\bf Z}\), and subtract from them their respective components along the $ j^{\rm th}_1 $ row direction,} yielding a modified $\mbf{Z}$.

\item Next, of all the so far unselected rows of the modified $\mbf{Z}$, we select the one with the largest 2-norm; we record its row number
in the original $\mbf{Z}$, and call it $j_2$.
We normalize the row and subtract its component along the still-remaining rows. This yields a further modified $\mbf{Z}$.

\item To clarify, we now have two indices selected ($j_1$ and $j_2$); and $n_{\trm{g}}n_{\trm{e}}-2$ rows of $\mbf{Z}$ remaining in contention, where for each of these remaining rows their components along the already-selected rows have been removed.

\item Proceeding as above, we select a third row (largest 2-norm among the $n_{\trm{g}}n_{\trm{e}}-2$ rows). We record its row number (call it $j_3$), normalize the row, and remove its component from the remaining $n_{\trm{g}}n_{\trm{e}}-3$ rows.

\item We proceed like this for as many rows as we wish to include in the reduced order model. A termination criterion based on the norm of the remaining part can be used if we wish.

\item In the end, we are interested only in the selected indices $j_1, j_2, \cdots, j_m$. What remains of the matrix $\mbf{Z}$ is discarded.

{\item In the above description, we could have removed the rows from \({\bf Z}\) after selecting them. That would leave a remaining matrix \({\bf Z}\) with progressively fewer rows as the calculation proceeded. That alternative algorithm is outlined below for ease of understanding.}
\end{enumerate}

{For intuitive understanding through a simple example, we now describe the above procedure using small hypothetical numbers. The data matrix has, say, $10$ rows (for simplicity). Let the 4$^{\rm th}$ row have the largest 2-norm. Let this row be denoted by $r_4$. We select the fourth row, remove it from the data matrix, and obtain a new matrix with 9 rows. We divide $r_4$ by its 2-norm to obtain a unit vector, say $e_4$. We take the inner product of $e_4$ with each of the  9 remaining rows, and subtract from these rows their components along $e_4$. The new data matrix has rows that are all orthogonal to $e_4$. Let the 5$^{\rm th}$ row of this remaining matrix have the largest 2-norm. We must remember that this index, namely 5, was actually 6 in the original 10-row data matrix. With that caveat, we are back to the second line of this paragraph, only with 9 rows instead of 10. We proceed in this way for as many rows as we wish to select, or until the 2-norms of the remining rows become small enough.}
	
\subsubsection{Finding matrix P}\label{P_mat}
With the indices chosen, a submatrix $\mbf{Z}_{\trm{s}}$ is assembled using the $(j_1,j_2,\dots,j_m)^{\trm{th}}$ rows ofthe original (and not the reduced) $\mbf{Z}$. We now solve the straightforward least squares problem,
    \begin{equation}\label{least_sqr_problem}
		\mbf{P} = \argmin_{\hat{\mbf{P}}\in \mathbb{R}^{r\times m}}\;\;{\lvert\lvert \mbf{R}^{\top}\mbf{A}\mbf{Z}- \hat{\mbf{P}}\mbf
			{Z}_{\trm{s}} \rvert\rvert}_{\trm{F}},	
	\end{equation}
where \(\lvert\lvert\cdot\rvert\rvert_{\trm{F}}\) denotes the Frobenius norm.
The above problem, upon taking matrix transposes, is routine in, e.g., Matlab.
The final reduced order model becomes
	\begin{subequations}
		\begin{equation}\label{ROM}
			\tilde{\mbf{M}}\ddot{\bsy{\xi}}+\tilde{\mbf{K}}\bsy{\xi}+\mbf{P}\bsy{z}_{\trm{s}}=\mbf{R}^\top\bsy{f}_0(t).
		\end{equation}
		\begin{multline}\label{ROM_z}
			\dot{\bsy{z}}_{\trm{s}}=\left(\bar{A}-\alpha\: \trm{sign}(\dot{\bsy{\chi}}_{\trm{s}}\circ\bsy{z}_{\trm{s}})\circ\abs{\bsy{z}_{\trm{s}}}^{\circ\, n_{\trm{h}}}-\beta \abs{\bsy{z}_{\trm{s}}}^{\circ\, n_{\trm{h}}}\right)\circ\dot{\bsy{\chi}}_{\trm{s}}, 
			\\ \dot{ \bsy{\chi}}_{\rm s}=\mbf{B}_{\rm s}\mbf{R}\dot{\bsy{\xi}}, 
		\end{multline}
	\end{subequations}
where $ \mbf{B}_{\rm s} $ is a submatrix of $ \mbf{B} $ constructed by retaining the $(j_1,j_2,\dots,j_m)^{\trm{th}}$ rows of the latter.

Equations \ref{ROM} and \ref{ROM_z} are equivalent to a $2r+m$ dimensional system of first order equations ($r$ modes and $m$ hysteretic variables). Note that the right hand side
of Eq.\ (\ref{ROM}) may seem like it has a large number of elements, but in practice for many problems, external forcing is restricted to a few locations or can be reduced to a few effective locations (e.g., if the forces acting at several nodes maintain fixed proportions to each other). We do not investigate this
aspect of size reduction because we have so far not made simplifying assumptions about $\bsy{f}_0(t)$. When the forcing is zero, of course, the right hand side is zero.
		
We now turn to the results obtained using, first, our integration routine; and then our approach for developing reduced order models.

\section{Results}\label{results}

Our main aim is accurate simulation of hysteretic dissipation, which is most easily seen in the unforced decaying response of the structure. So we will first consider some unforced transient responses of the structure to examine both stability and numerical accuracy of our semi-implicit integration algorithm. Subsequently we will present reduced order models and show their efficacy compared to the full model. 

In the results from direct numerical integration using our proposed semi-implicit algorithm, we will check for three things: (i) the theoretically expected power law decay for small amplitude vibrations, (ii) the absence of instabilities arising from higher modes, and (iii) overall accuracy.

For error calculations with low dimensional systems, i.e., when the number of elements in the FE model is small, we will use Matlab's \texttt{ode15s} for comparison with both the absolute and relative error tolerances set to $10^{-10}$; this is because \texttt{ode15s} has built-in adaptive step sizing and is accurate when it works. For high dimensional systems \texttt{ode15s} does not work and we will use the proposed semi-implicit algorithm itself with a very small step size $\left( h=2^{-23} \right)$ for error estimates.

We now select some parameter values to be used in the simulation.  

\subsection{Choice of parameter values}\label{parameter_choose}
We consider a cantilever beam with {Young's modulus \(E=200\) GPa and density \(\rho=7850\, \trm{kg}\,{\rm m}^{-3}\); of length $1 \,\trm{m}$ and square cross section of \(2\, {\rm cm}\times 2\, {\rm cm}\)}. This yields a flexural rigidity $EI = 2666.7\,\trm{N}\trm{m}^2$ and mass per unit length $ \bar{m} = 3.14\, \trm{kg}\,\trm{m}^{-1}$. 

For many metals, 0.2\% strain is near the border of elastic behavior. For the beam parameters above, if the beam is statically deflected under a single transverse tip load, then 0.2\% strain at the fixed end of the beam corresponds to a tip deflection of 6 cm, which is taken as a reasonable upper bound for the vibration amplitudes to be considered in our simulations below.

Next, we consider the hysteretic damping itself. The index $n_{\rm h}$ in the Bouc-Wen model is primarily taken to be 0.5 for reasons explained a little later. A larger value is considered in subsection \ref{Lnh}, and only in subsection \ref{Lnh}, to clarify an issue in convergence.
The parameters $\alpha$ and $\beta$ are somewhat arbitrary; we have found that the values $\alpha=0.8$ and $\beta=0.5$ yield hysteresis loops of reasonable shape. It remains to choose the Bouc-Wen parameter $\bar A$.

It is known that for small amplitude oscillations, the Bouc-Wen dissipation follows a power law. An estimate for the upper limit of forcing amplitude in the Bouc-Wen model, below which the power law should hold, is given in \cite{bhattacharjee2013dissipation} as
\begin{equation}\label{Abar_val1}
	    \chi_{\rm max}= \left( 2\,{\frac {\bar{A}^{2-2\,n_{\trm{h}}} \left( 1+n_{\trm{h}} \right) 
 \left( 1+2\,n_{\trm{h}} \right)  \left( 2+3\,n_{\trm{h}} \right) }{ \left( 2\,{n_{\trm{h}}}^{2}{
\alpha}^{2}+4\,n_{\trm{h}}{\alpha}^{2}-n_{\trm{h}}{\beta}^{2}+2\,{\alpha}^{2} \right) 
 \left( 2+n_{\trm{h}} \right) }} \right) ^{{\frac {1}{2\,n_{\trm{h}}}}}.
\end{equation}

Choosing $\chi_{\rm max}$ to correspond to the abovementioned 0.2\% strain at the fixed end, in turn corresponding to a static free-end deflection of 6 cm for the above beam, we find from Eq.\ (\ref{Abar_val1}) that
$$\bar A = 0.065.$$

\begin{figure}[h!]
	\centering
	\includegraphics[scale=0.5]{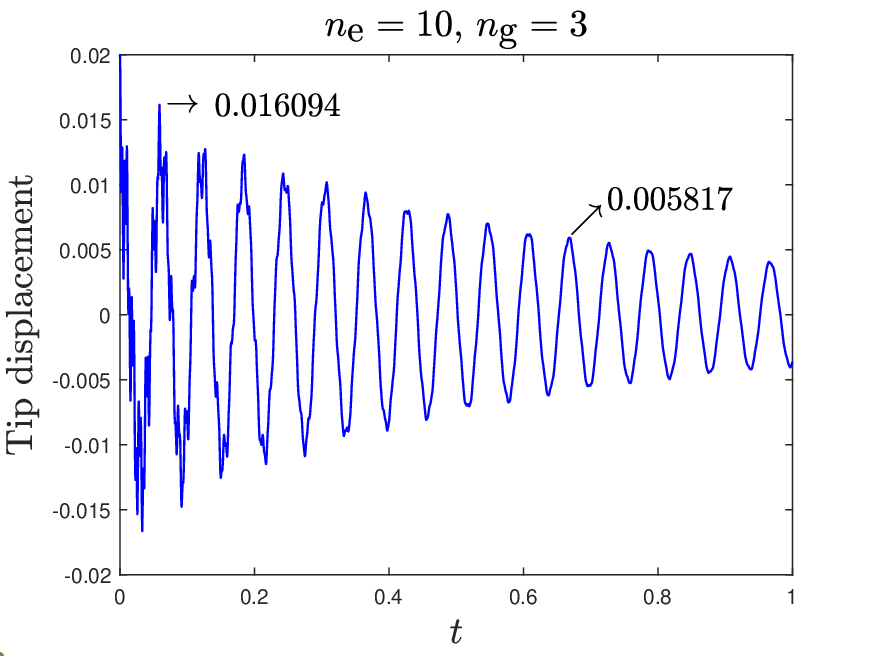}
	\caption{Tip displacement for $\gamma_{\trm{h}}=3000$ and $ n_{\rm h}=0.5 $ shows an approximate equivalent damping of about
1.6\%, as per Eq.\ (\ref{zeta_est}). Here 10 beam elements were used, with 3 hysteresis Gauss points per element. Time integration was done
using Matlab's \texttt{ode15s} with error tolerances set to $10^{-10}$.}
	\label{decay}
\end{figure}

Only one physical model parameter remains to be chosen, namely $\gamma_{\trm{h}}$. To choose this parameter value, we note that the amplitude decay in free vibrations with Bouc-Wen hysteretic dissipation is not exponential. Thus, the idea of percentage of damping is only approximately applicable. Upon looking at the reduction in amplitude after a certain number of oscillations (say $M$), an equivalent ``percentage damping'' can be approximated using the formula
\begin{equation}\label{zeta_est}
	  \zeta_{\trm{equiv}}\approx\frac{1}{2\pi M}\ln\left(\frac{A_1}{A_{1+M}}\right).
\end{equation} 
Metal structures often have 1-2\,\% damping \cite{orban2011damping}. We will choose values of the hysteretic dissipation $\gamma_{\trm{h}}$ (recall Eq.\ (\ref{eq_gov})) to obtain $0.01 \le \zeta_{\trm{equiv}} \le 0.02$.
Numerical trial and error show that
$$\gamma_{\trm{h}}=3000 \, {\rm Nm}$$
is a suitable value (see figure \ref{decay}).

\subsection{Semi-implicit integration: stability and accuracy}

Before we examine the performance of our proposed semi-implicit integration method,
we will verify that the FE model indeed displays power law damping at small amplitudes, as predicted by theoretical analyses of the Bouc-Wen model.

The small amplitude dissipation per cycle of the Bouc-Wen model is proportional to amplitude to the power $n_{\trm{h}}+2$ \cite{bhattacharjee2013dissipation}. Since the energy in the oscillation is proportional to amplitude squared, we should eventually observe
small amplitude oscillations in the most weakly damped mode with amplitude $A$ obeying
$$A\dot{A}=\mcl{O}\left(A^{n_{\trm{h}}+2}\right)$$
whence for $n_{\trm{h}} > 0$
\begin{equation}
\label{power_law_dissipation}
  A=\mcl{O}\left(t^{-\frac{1}{n_{\trm{h}}}}\right).
\end{equation}

For the Bouc-Wen model, letting $n_{\trm{h}} \rightarrow 0$ produces hysteresis loops of parallelogram-like shape, and so we prefer somewhat larger values of $n_{\trm{h}}$; however, to have a significant decay rate even at small amplitudes, we prefer somewhat smaller values of $n_{\trm{h}}$.
As a tradeoff, we have chosen $n_{\trm{h}}=\half$. We expect an eventual decay of vibration amplitudes like $1/t^2$.

\begin{figure}[h]
	\centering
	\includegraphics[width=\linewidth]{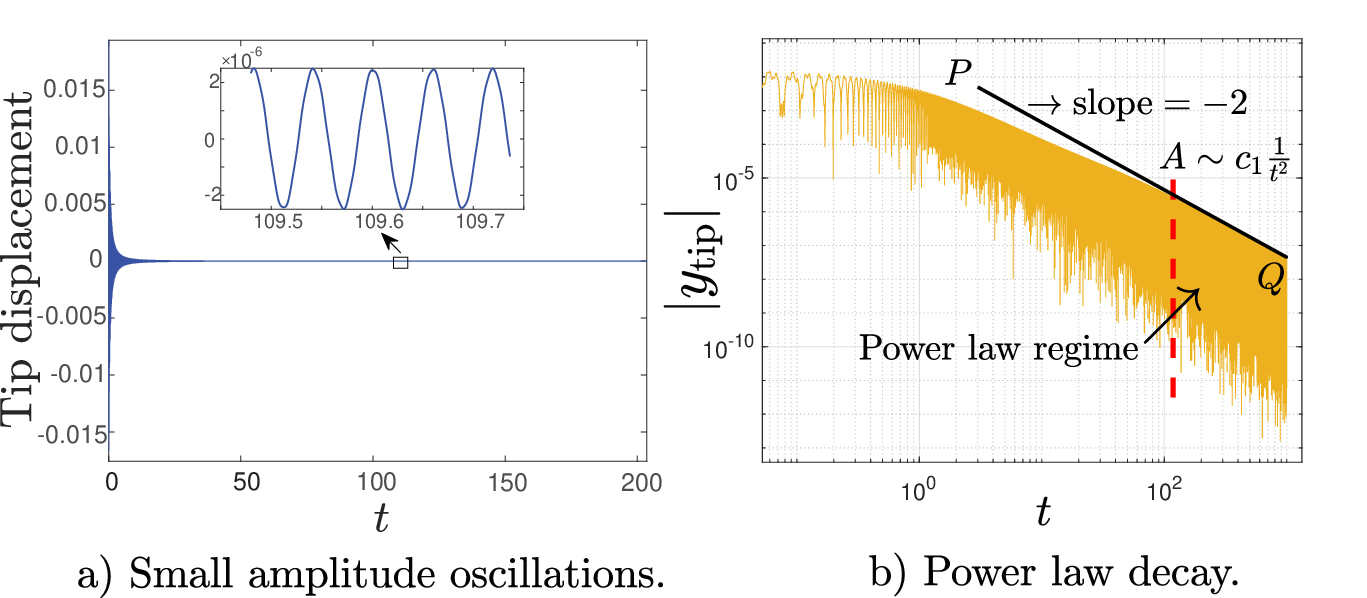}
	\caption{The long term oscillation in the beam tip response shows very slow power law decay when $ n_{\rm h}=\half$. A small portion of the solution is shown zoomed within the left subplot. The aim of this simulation is to show that the model and numerical integration method together retain the correct asymptotic
behavior far into the small-amplitude regime.}
	\label{beam_decay_image}
\end{figure}

For our beam model, with 10 elements and 3 Gauss points per element, and with our semi-implicit numerical integration algorithm\footnote{%
	Matlab's \texttt{ode15s} struggles with such long simulations on a small desktop computer; our algorithm works quickly.}, the computed tip displacement asymptotically displays the expected power law decay rate, as seen on linear axes in Fig.\ (\ref{beam_decay_image}a) 
and more clearly in the logarithmic plot of Fig.\ (\ref{beam_decay_image}b). The frequency of the residual oscillation is close to the first undamped natural frequency of the beam. Higher modes are not seen in the long-term power law decay regime.

\begin{figure}[h]
	\centering
	\includegraphics[width=\linewidth]{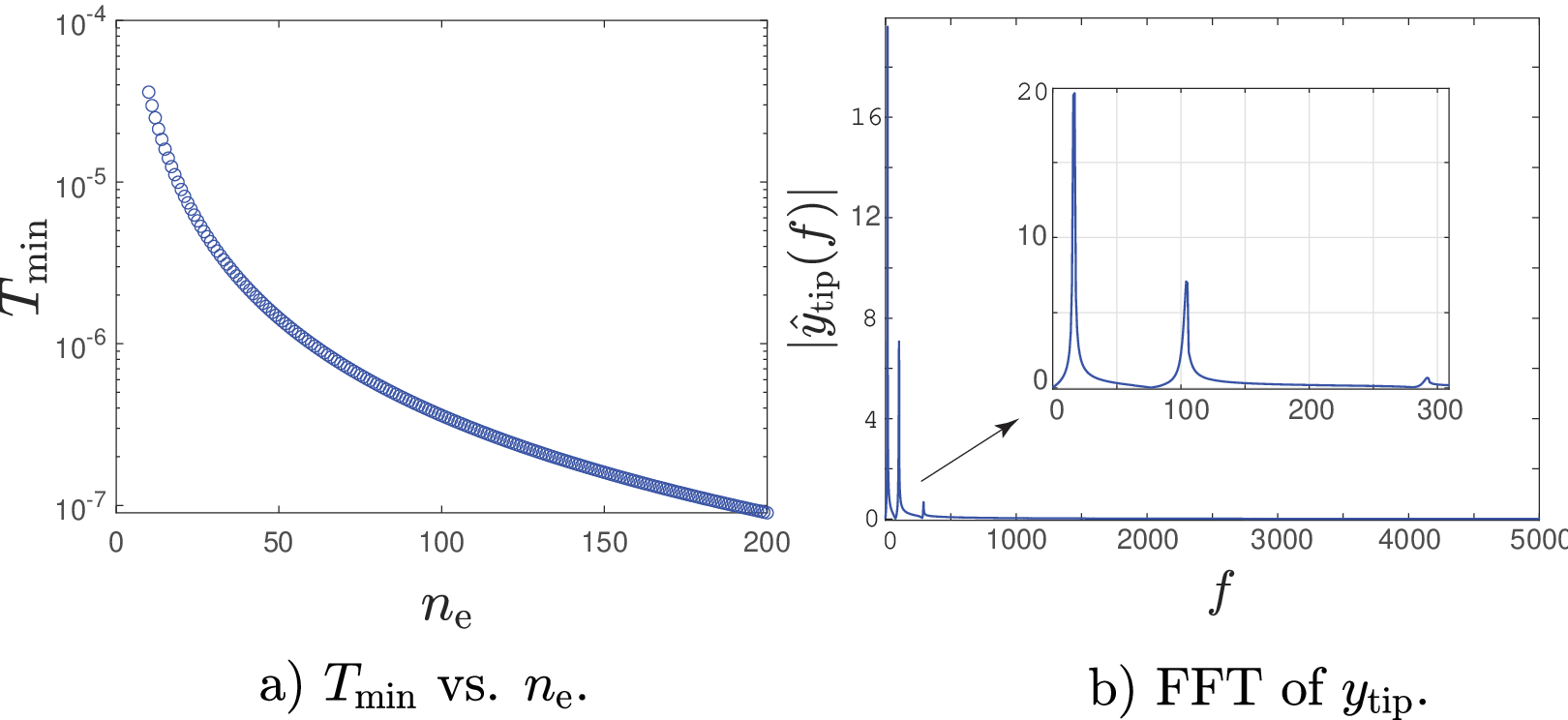}
	\caption{(a) Variation of $ T_{\min} $ with $ n_{\trm{e}} $ and (b) Frequency content of the transient tip displacement response when the first three modes are disturbed in a uniform FE model with 100 elements with $ n_{\rm h}=0.5$.}
    \label{T_min_and_ytip_fft}
\end{figure}

Having checked the asymptotic power law decay rate, we next ensure that the semi-implicit algorithm does not produce spurious oscillations in higher modes within the computed solution. This issue is important because, 
with increasing mesh refinement in the FE model, very high frequencies are unavoidable. While those high frequency modes exist in principle, their response should be small if any external excitation is at low frequencies and initial conditions involve only the lower modes.
To examine this aspect, we denote the time period of the highest mode present in the structural model by $T_{\trm{min}}$. As the number of elements increases, $T_{\trm{min}}$ decreases, as indicated in
Fig.\ (\ref{T_min_and_ytip_fft}a). 

We now use our semi-implicit method to simulate an FE model with 100 elements and with nonzero initial conditions along only the first 3
modes. In the computed solution, we expect to see frequency content corresponding to the first three modes only. Figure \ref{T_min_and_ytip_fft}b indeed shows only three peaks. Spurious responses of the higher modes are not excited. Note that this simulation was done with a time step of $h=10^{-4}$, which is more than 275 times larger than the time period of the highest mode for $n_{\rm e} = 100$, which is 
$ T_{\min}=3.6\ts 10^{-7}\,\trm{sec} $ (see Fig.\ (\ref{T_min_and_ytip_fft}a)). It is the implicit part of the code that allows stable numerical integration with such a large step size.

Having checked that spurious oscillations in very high modes are {\em not} excited, it remains to check that accurate results are obtained with time steps $h$ which are sufficiently small compared to the highest mode of interest. To this end, the convergence of the solution with decreasing $h$ is depicted
in Fig.\ (\ref{match_1}). For the beam modeled using 10 elements, the first five natural frequencies are
$$16.3,  \, 102.2, \, 286.2,\, 561.3 \mbox{ and } 929.3 \mbox{ Hz}.$$
Numerical simulation results using our semi-implicit algorithm are shown in Fig.\ (\ref{match_1}) for $h = 10^{-3}$, $10^{-4}$ and $10^{-5}$. The overall solution is plotted on the left, and a small portion is shown enlarged on the right. Only 10 elements were used for this simulation to allow use of Matlab's {\tt ode15s}, which has adaptive step sizing and allows error tolerance to be specified (here we used $10^{-10}$). It is seen in the right subplot that although all three solutions from the semi-implicit method are stable, the one with $h=10^{-3}$ does not do very well in terms of accuracy; the one with $h=10^{-4}$ is reasonably close and may be useful for practical purposes; and the one with
$h=10^{-5}$ is indistinguishable at plotting accuracy from the {\tt ode15s} solution. The match between our semi-implicit integration with 
$h=10^{-5}$ and Matlab's {\tt ode15s} with error tolerances set to $10^{-10}$ indicates that both these solutions are highly accurate.

\subsection{Order of convergence}
\label{sscn_er_conv}

Using simulations with relatively few elements and over relatively short times, we can compare the results obtained from our semi-implicit method (SIM) with \texttt{ode15s}, and examine convergence as $h$ is made smaller.
\begin{figure}[h!]
    \centering
    \includegraphics[width=\linewidth]{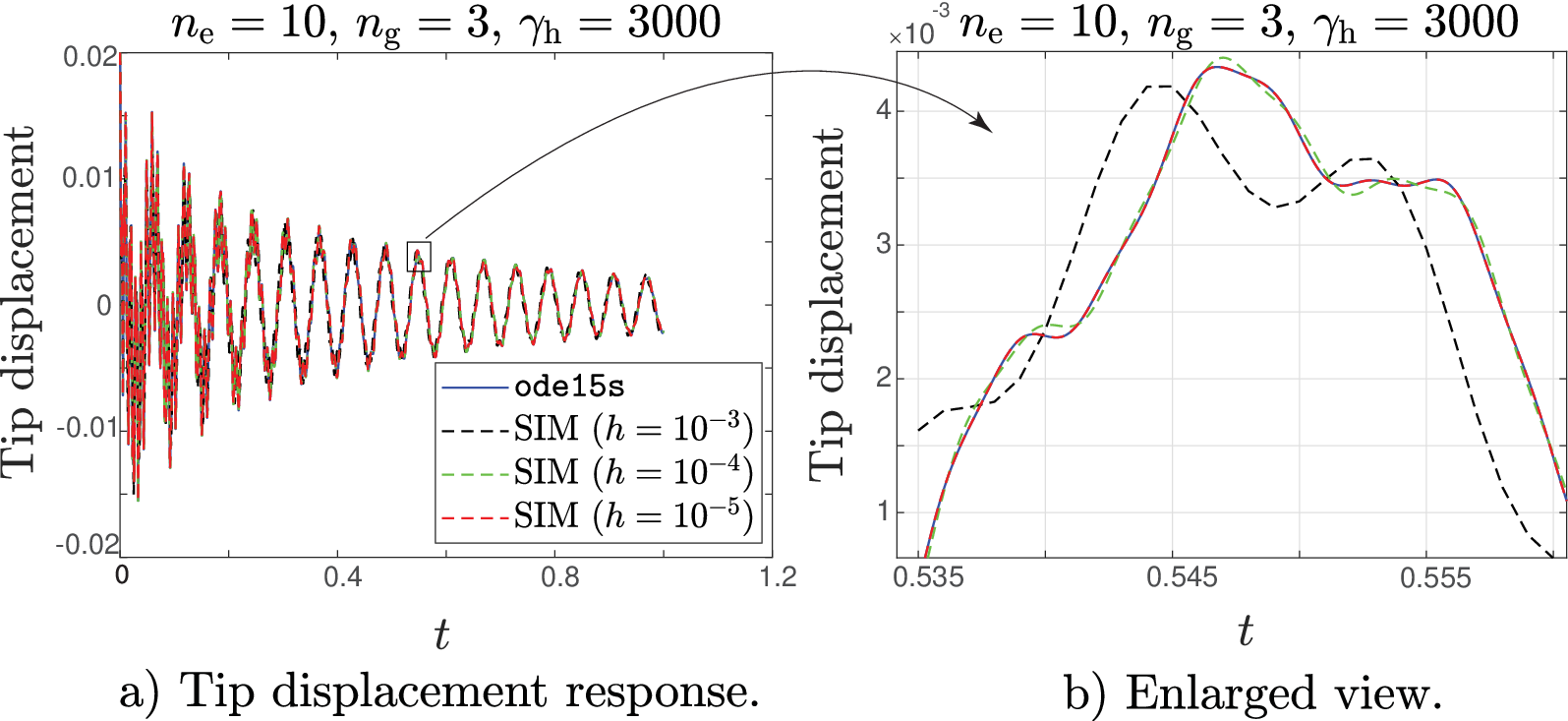}
    \caption{Tip displacement response calculated using different time steps (compared to \texttt{ode15s}) with $ n_{\rm h}=0.5$.}
    \label{match_1}
\end{figure}
Figure \ref{match_1} shows the tip response for an FE model with 10 elements for different time step sizes and compares them with the \texttt{ode15s} solution. 
Here we will use two different error measures.

\subsubsection{RMS error}

We choose a fairly large number ($N_\trm{E} = 128$) of points evenly spaced in time, and for each $h$ we calculate the response at those instants of time. The overall time interval was chosen to be [0,1]. Clearly, $h$ can be successively halved in a sequence of simulations to ensure that the solution is always
computed at exactly the $N_\trm{E}$ time instants of interest (along with other intermediate points).

\begin{figure}[h]
	\centering
	\includegraphics[width=\linewidth]{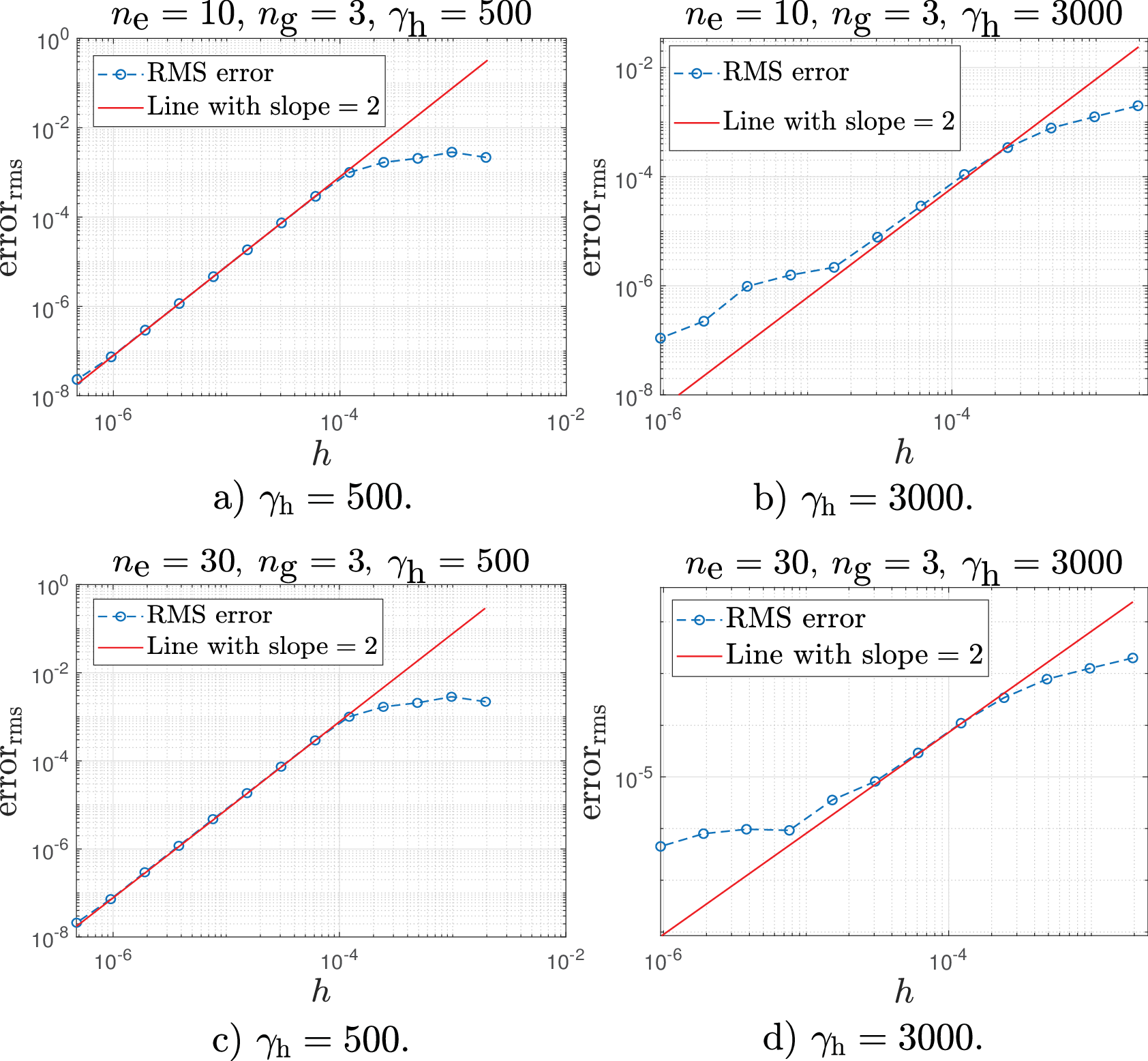}
	\caption{Time step vs.\ RMS error (128 equispaced points in time) for $ n_{\rm e}=10$ and 30 with $ n_{\rm h}=0.5 $.}
	\label{error_plot1}
\end{figure}

We define our RMS error measure as
\begin{equation}
	e_{\trm{rms}}(h)=\sqrt{ \frac{1}{N_{\trm{E}}}\sum_{k=1}^{N_{\trm{E}}}{\left(y_h(t_k)-y_{\trm{accurate}}(t_k)\right)^2}}, \quad
N_\trm{E} = 128.
\end{equation}
In the above, when the number of elements is modest (e.g., $n_{\rm e} = 10$), we use the highly accurate solution obtained from \texttt{ode15s} as  the
``accurate'' one.
With larger numbers of elements, \texttt{ode15s} cannot be used for validation. Having seen the overall accuracy of the semi-implicit method (SIM)
with fewer elements when compared with \texttt{ode15s}, we use the SIM solution with extremely small time steps as the ``accurate'' one for
error estimation with larger number of elements (e.g., $n_{\rm e} = 30$). 
Results are shown in Fig.\ (\ref{error_plot1}). It is seen that for relatively smaller values of $\gamma_{\trm{h}}$, a significant regime of
 approximately quadratic convergence is obtained (Fig.\ (\ref{error_plot1}a,\ref{error_plot1}c)). It means that for lightly damped structures the performance of the semi-implicit algorithm is excellent. For larger values of $\gamma_{\trm{h}}$, however, the convergence plot is more complicated, and there is no significant regime of quadratic convergence (Fig.\ (\ref{error_plot1}b,\ref{error_plot1}d)).
This is because the damping model is strongly non-analytic, and strong damping makes the role of that non-analyticity stronger. However, over a significant regime and in an average sense, it appears that the convergence is superlinear (i.e., the average slope exceeds unity in a loglog plot), and so the integration algorithm still performs well. A larger value will be considered in subsection \ref{Lnh}.

\begin{figure}[h!]
	\centering
	\includegraphics[scale=0.5]{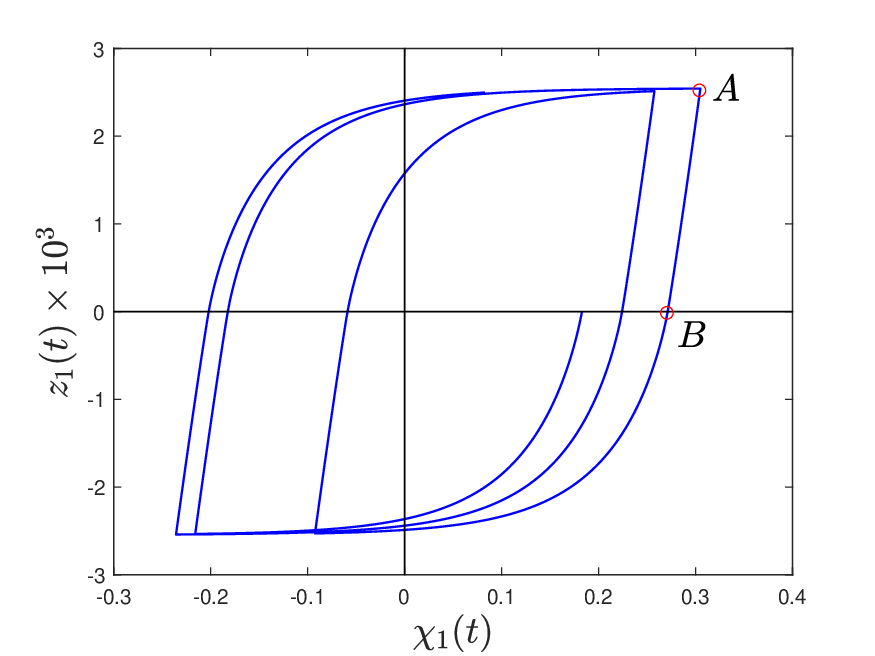}
	\caption{Hysteresis loop for the $z_1$ driven by $\chi_1$ (at Gauss point ``1'' near the fixed end).}
	\label{hys_loop}
\end{figure}

The role of nonanalyticity in the Bouc-Wen model, and the way in which it is handled in our semi-implicit algorithm, are worth emphasizing.
See Fig.\ (\ref{hys_loop}).
In the figure, point A shows a discontinuous slope change due to the sign change of the driving term,
$\dot{\chi}$. This point is handled with an {\em explicit} attempt to locate the instant of change in direction, with one integration step taken on each side of that instant. Point B indicates a sign change in $z_1$. Our proposed algorithm does not separately identify sign changes of $z$ within one time step, in the interest of simplicity.
Due to nonanalyticity at both points A and B, integration errors increase when the nonsmoothness dominates (high $\gamma_{\trm{h}}$ and/or small $n_{\rm h}$). A larger value will be considered in subsection \ref{Lnh}.

\subsubsection{Error at a fixed instant of time}

We now consider the absolute value of error at some fixed instant of time $t=\tau$,
\begin{equation}
	e_{\tau}(h)=\lvert y_h(\tau)-y_{\trm{accurate}}(\tau)\rvert.
\end{equation}
Here, we use $\tau = 1$.

\begin{figure}[h!]
	\centering
	\includegraphics[width=\linewidth]{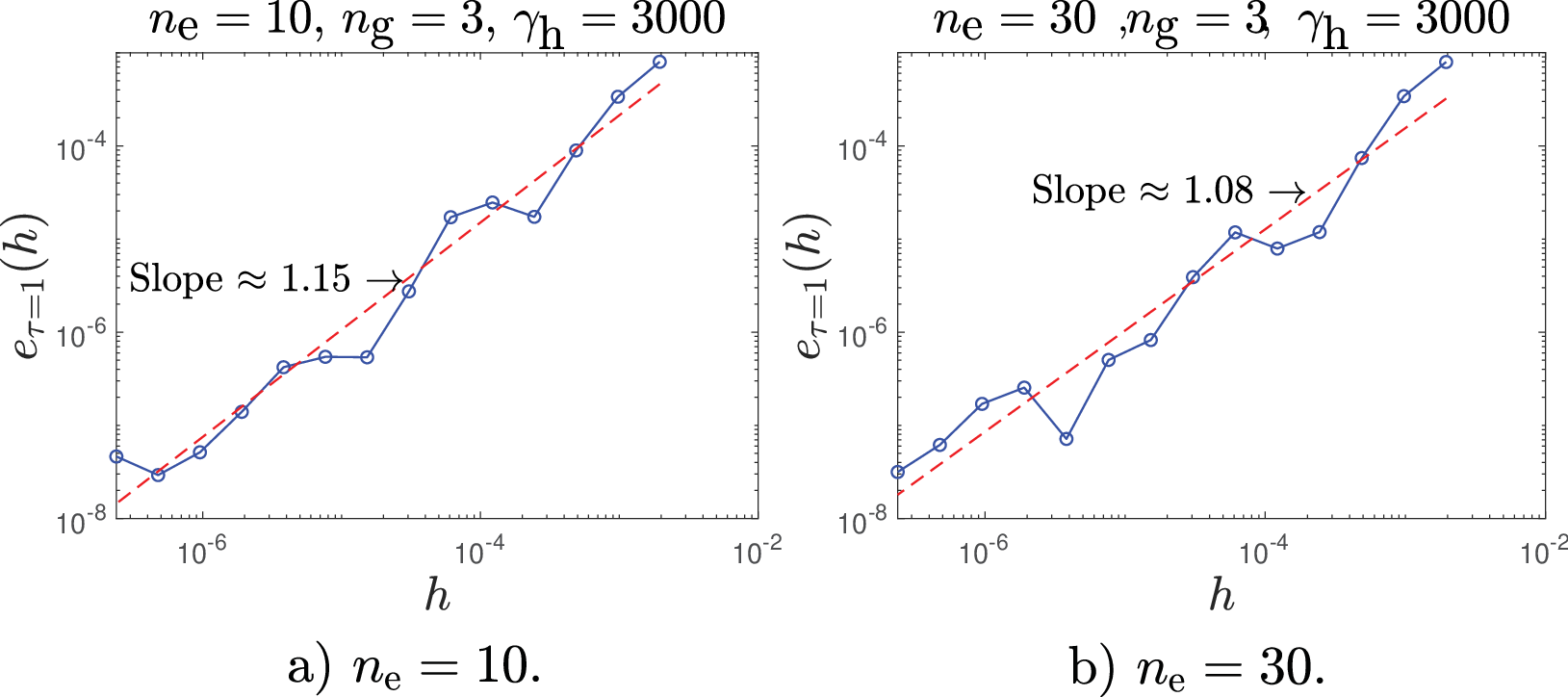}
	\caption{Error at $t=1$, for FE models with 10 and 30 elements with $n_{\rm h}=0.5$.}
	\label{fxt_error}
\end{figure}

Straight line fits on loglog plots of $ e_{\tau}(h) $ versus $ h $  in Fig.\ (\ref{fxt_error}a,\ref{fxt_error}b) show average slopes that slightly
exceed unity. While these slopes are not rigorously proved to exceed unity, the overall convergence rate of the integration algorithm may be considered
roughly linear (on average) over a sizeable range of step sizes. It is emphasized that these error computations are done in a regime where
(i) Matlab's 
\texttt{ode15s} does not work at all, (ii) explicit methods are unstable unless extremely small time steps are used, (iii) properly implicit algorithms are both complex and not guaranteed to converge, and (iv) $n_{\rm h} = 0.5$ in the Bouc-Wen model is relatively small. Considering these four difficulties, the semi-implicit method (SIM) proposed in this paper may be said to be simple, effective, and accurate.

\subsection{Larger $n_{\rm h}$}
\label{Lnh}
Everywhere in this paper except for this single subsection, we have used $n_{\rm h} = 0.5$. In this subsection only, we consider $n_{\rm h}=1.5$. Due to the greater smoothness of the hysteresis loop near the point B of Fig.\ (\ref{hys_loop}), we expect better convergence for this higher value of $n_{\rm h}$. 

Some parameter choices must be made again.
Using the yield criteria used in section \ref{parameter_choose}, for $ n_{\rm h}=1.5 $, we find we now require
$$\bar{A}=608.9\,.$$

Subsequently, we find an approximate equivalent damping ratio $\zeta_{\rm equiv}=0.015$  for $ \gamma_{\rm h}=0.3$.  It is interesting to note that with the change in $ n_{\rm h}$ and for the physical behavior regime of interest, $\bar{A}$ and $ \gamma_{\rm h}$ have individually changed a lot but their product has varied only slightly (195 Nm for about 1.6\% damping in the $n_{\rm h}=0.5$ case, and 183 Nm for about 1.5\% damping
 in the $n_{\rm h}=1.5$ case).

\begin{figure}[h]
	\centering
	\includegraphics[scale=0.5]{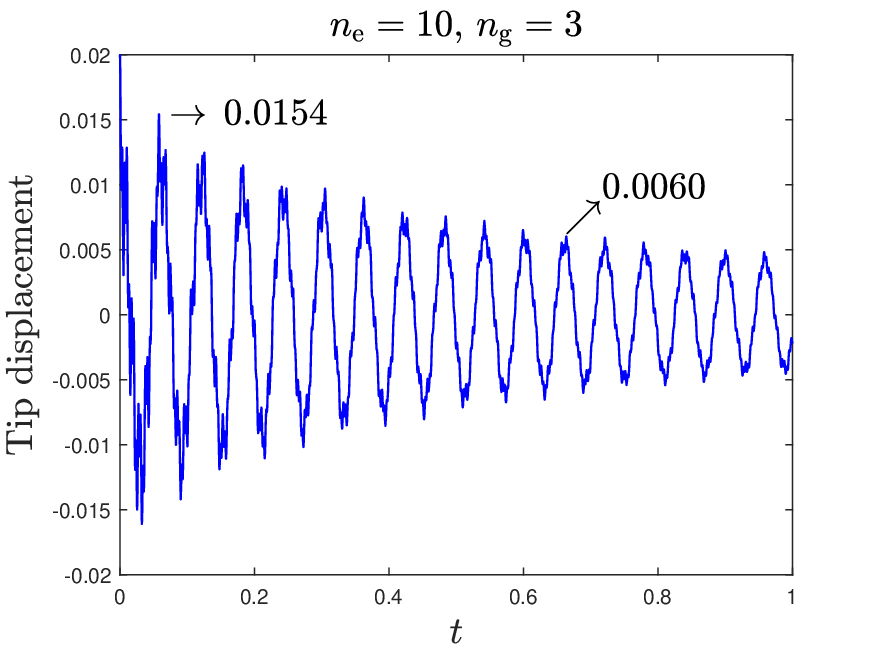}
	\caption{Tip displacement for $\gamma_{\trm{h}}=0.3$ and $ n_{\rm h}=1.5 $ shows an approximate equivalent damping of about
		1.5\%, as per Eq.\ (\ref{zeta_est}). Here 10 beam elements were used, with 3 hysteresis Gauss points per element. Time integration was done
		using Matlab's \texttt{ode15s} with error tolerances set to $10^{-10}$.}
	\label{decay_nh_1point5}
\end{figure}

The decay of tip response amplitude (Fig.\ (\ref{decay_nh_1point5})) for these parameters (with $n_{\rm h}=1.5$)
looks similar to the case studied in the rest of this paper ($n_{\rm h}=0.5 $, with attention to Fig.\ (\ref{decay})).

\begin{figure}[h!]
	\centering
	\includegraphics[width=\linewidth]{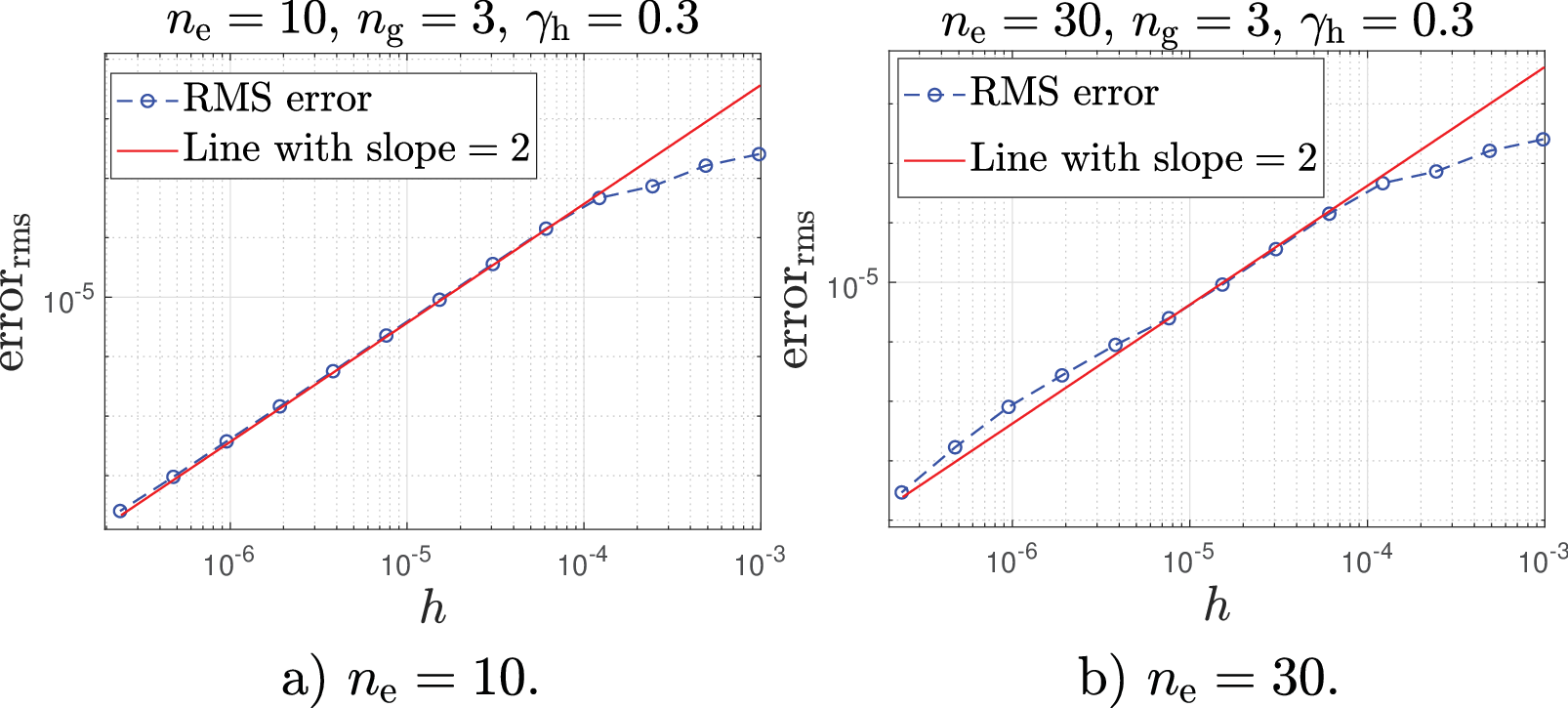}
	\caption{Time step vs.\ RMS error (128 equispaced points in time) for $ n_{\rm e}=10$ and 30 with $ n_{\rm h}=1.5 $. The linear fit for 10 elements has slope 2.0. The linear fit for 30 elements has slope 1.9; another line with slope 2 is shown for comparison.}
	\label{error_plot3}
\end{figure}
The main point to be noted in this subsection is that with $n_{\rm h}=1.5$, $\bar{A}=608.9$, $\gamma_{\rm h}=0.3$, and all other parameters the same as before,
we do indeed observe superior convergence over a significant range of step sizes: see Fig.\ (\ref{error_plot3}) for FE models with 10 and 30 elements,  and compare with Fig.\ (\ref{error_plot1}). For the case with 10 elements, the convergence is essentially quadratic. With 30 elements the convergence is slightly slower than quadratic, but much faster than linear. Note that these estimates are from numerics only: analytical estimates are not available.

As mentioned above some of the difficulty with accurate integration of hysteretically damped structures comes from the zero crossings of the hysteretic variable $z$ itself. In this paper, for simplicity, we have avoided special treatment of these zero crossings. However, the results of this subsection show that the effect of these zero crossings is milder if $n_{\rm h}$ is larger.
 
We now turn to using results obtained from this semi-implicit integration method to develop data-driven lower order models of the hysteretically damped structure.

\subsection{Reduced order models}
In our beam system, we have 2 degrees of freedom per node (one displacement and one rotation). Let the total number of differential equations being solved, in first order form, be called $N_{\rm D}$. If there are $n_{\rm e}$ elements, we have
$N_{\rm D} = 4 n_{\rm e} + n_{\rm e}n_{\rm g}$ first order differential equations. In a reduced order model with 
$r$ modes and $m$ Gauss points, we have $N_{\rm D}=2r+m$ first order differential equations. Although the size of the problem is reduced significantly in this way, the accuracy can be acceptable.

For demonstration, we consider two FE models of a cantilever beam, with hysteresis, as follows:
\begin{itemize}
	\item[(i)] 100 elements with 3 Gauss points each ($N_{\rm D} = 700$).
	\item[(ii)] 150 elements with 3 Gauss points each  ($N_{\rm D} = 1050$).
\end{itemize}

For each of the two systems above, the datasets used for selecting the subset of hysteretic states (or Gauss points) were generated by solving the full systems 60 times each for the time interval 0 to 1 with random initial conditions and time step $h=10^{-4}$, exciting only the first 3 modes ($r=3$). Data from each of these 60 solutions were retained at 1000 equispaced points in time ($N_t=1000$; recall subsection \ref{z_selection}).

\begin{figure}[h!]
	\centering
	\includegraphics[width=\linewidth]{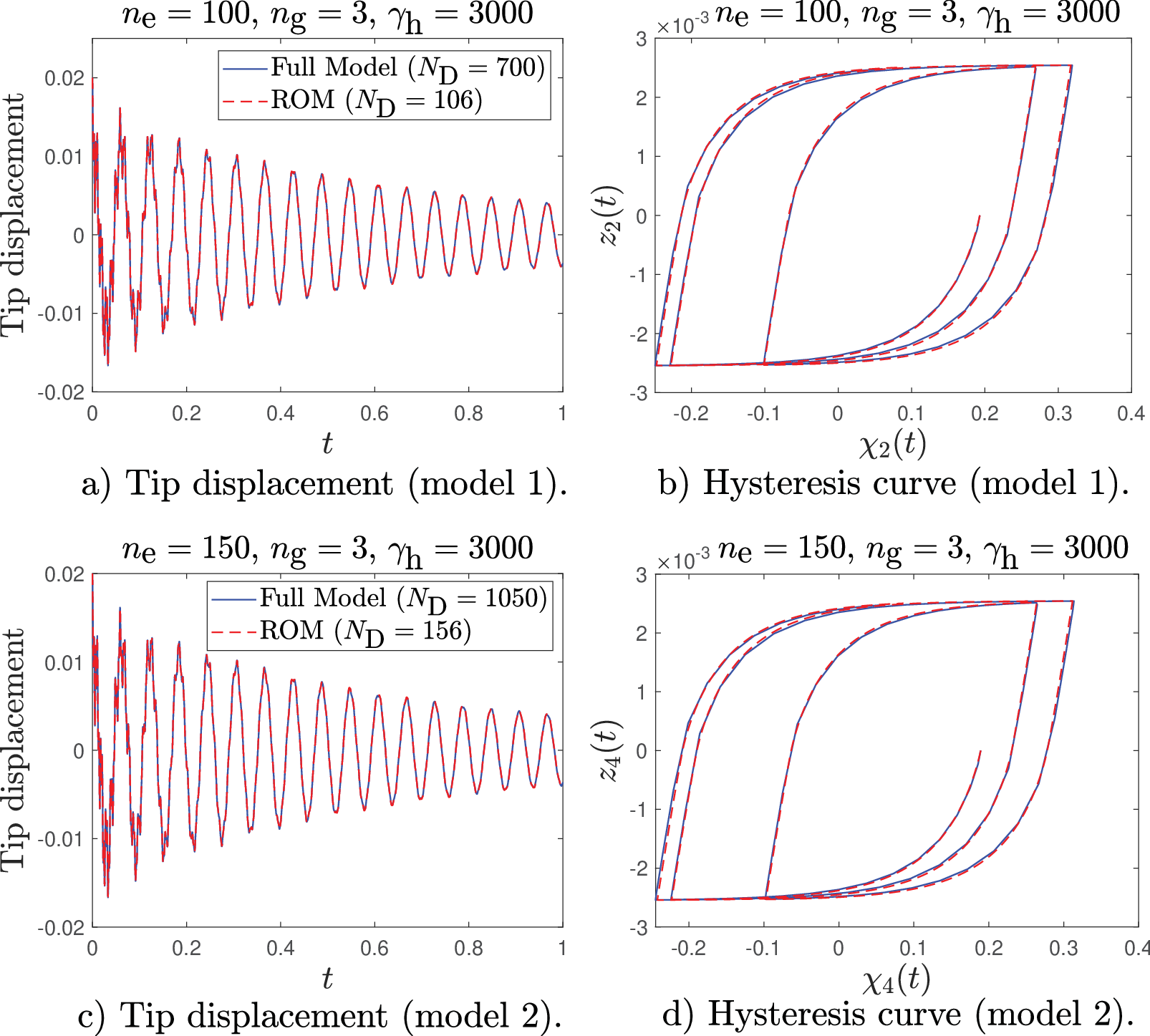}
	\caption{Comparison between ROM and full model, for two cases (see text for details). Tip displacements (left) and hysteresis curve (right).
{In subplots (b) and (d), the retained hysteresis Gauss point closest to the fixed end of the beam has been selected for display.}}
	\label{rom_response_fig}
\end{figure}

All reduced order model (ROM) results below are for $r=3$ retained modes. Further, the number of hysteresis Gauss points is denoted by $m$.
Results for $m=n_{\rm e} $ are shown in Fig.\ (\ref{rom_response_fig}a,\ref{rom_response_fig}c).

We now quantitatively assess the accuracy of the ROMs. To this end, we write $y^{(m)}_{\rm tip(ROM)} (t)$ and $y_{\rm tip(FM)}(t)$ for the ROM and full model outputs respectively. 
We then compute the error measure
\begin{equation}
	\mcl{E}_{\rm rms}=\sqrt{ \frac{1}{N_{\rm E}}\sum_{k=1}^{N_{\rm E}}  {\left(y^{(m)}_{\rm tip(ROM)}(t_k)-   y_{\rm tip(FM)}(t_k) \right )^2}  }, \, N_{\rm E}=10001,
\end{equation} 
for different values of $m$ and for an integration time interval $[0,1]$  (with $h=10^{-4}$).
\begin{figure}[h!] 
	\centering
	\includegraphics[width=\linewidth]{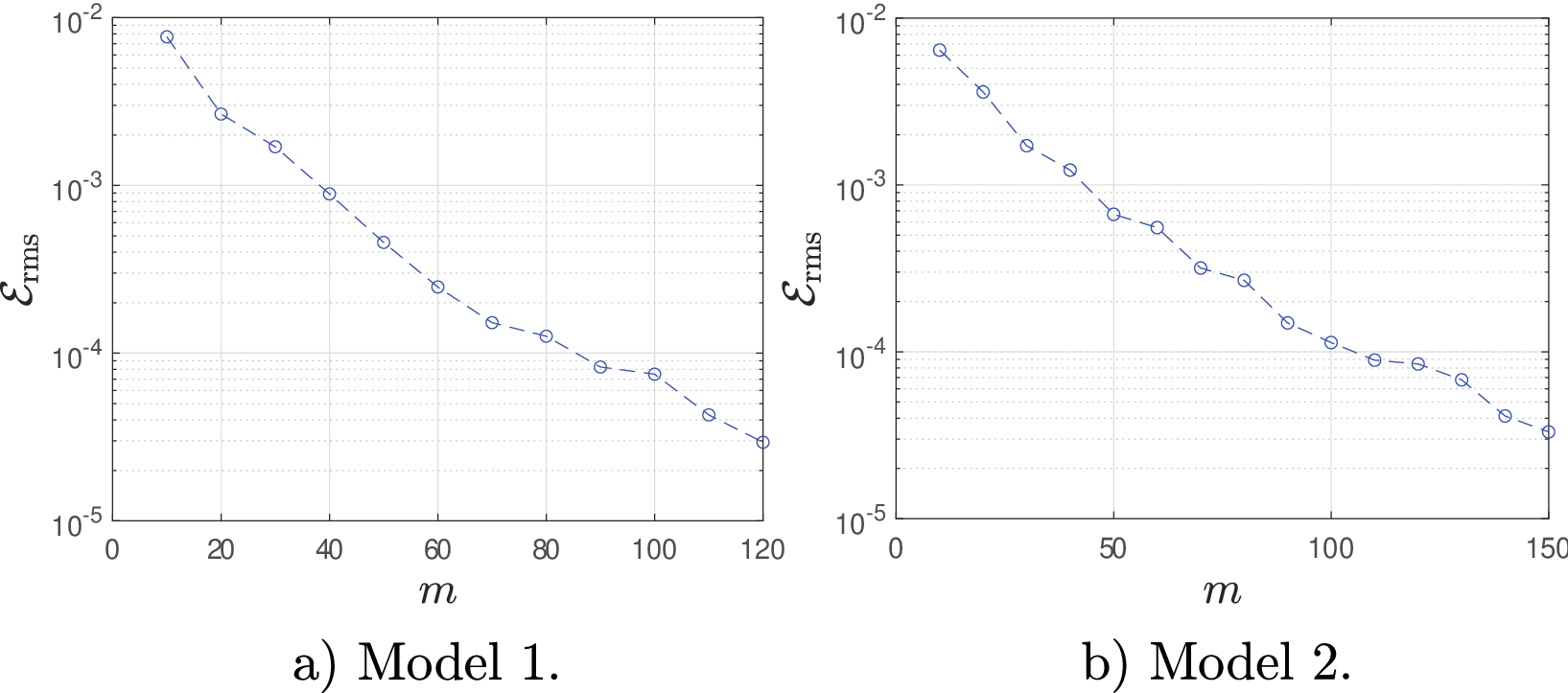}
	\caption{Error convergence of ROM with increasing number of Gauss points retained. For comparison, if $y^{(m)}_{\rm tip(ROM)}(t)$ is set identically to zero, the error measure obtained is 0.006 for both models. Thus, for reasonable accuracy, $m \ge 50$ may be needed.}
	\label{rom_error_fig} 
\end{figure}
The variation of $ \mcl{E}_{\rm rms} $ with $ m $, for both models, is shown in Fig.\ (\ref{rom_error_fig}). This error measure is not normalized. If we wish to normalize it, then we should use the quantity obtained when we set $y^{(m)}_{\rm tip(ROM)}(t)$ identically to zero: that quantity is 0.006 for both models. Thus, reasonable accuracy is obtained for $m \ge 50$, and good accuracy (about 1\% error) is obtained only for $m>100$.
These numbers refer to a specific case with random initial conditions; however, Fig.\ (\ref{rom_error_fig}) is representative for other, similar, initial conditions (details omitted).

{Finally, we report on run times needed on a modest laptop computer with the different levels of modeling described above. A representative initial conditions similar to the above was selected. The simulation duration was chosen to be 1 second, and the time step taken was $10^{-4}$ seconds. The following results were obtained after averaging run times for 4 runs each. (i) The full simulation with our algorithm took an average of 27.22 seconds per run. (ii) After projecting on to 3 normal modes but retaining all 450 hysteretic states, each run took an average of 1.3 seconds. (iii) Finally, with 3 normal modes and hysteretic states reduced from 450 to 150, each run took an average of 0.84 seconds. Note that we could save further
run time by taking larger time steps after model order reduction and/or retaining even fewer hsyteretic states.}

\section{Conclusions}
Structural damping is nonlinear and, empirically, dominated by rate-independent mechanisms. Hysteresis models are therefore suitable, but present numerical difficulties when used in large finite element models. Fully implicit numerical integration of such highly refined FE models presents difficulties with convergence in addition to algorithmic complexities. With this view, we have proposed simpler but effective approaches, in two stages, to simulations of such structural dynamics. The first stage consists of a semi-implicit integration routine that is relatively simple to implement, and appears to give linear convergence or better. Moreover, the time steps required for stability can be much larger than the time period of the highest mode in the FE model. Subsequently, we have used the results from that semi-implicit integration to develop data-driven reduced order models for subsequent rapid case studies or other simulations.

Thus, our contribution is twofold: first, we present a simple and practical numerical integration method that can be easily implemented; second, we use the results from that integration algorithm to further reduce the size of the model.

Although we have worked exclusively with a cantilever beam in this paper, our approach can be directly extended to frames that can be modeled using beam elements. We hope that future work will also examine ways in which the present approach can be extended to genuinely two- or three-dimensional structures. While we have not worked such cases out yet, we are hopeful that the issues addressed in the present paper will help to tackle such higher-dimensional problems.

\section*{Acknowledgements}
Aditya Sabale helped BG with the initial finite element formulation for beams. AC thanks the Department of Science and Technology, Government of India, for earlier support on a project to study hysteretic damping; this work arose from a continued interest in that topic.

\appendix

\section{Finite element formulation details}
\label{appB}

Starting with
\begin{equation}
	\hat{u}_{(\trm{e})}=\sum_{k=1}^{4} {\psi_{(\trm{e})}}_k(x){q_{(\trm{e})}}_k(t),
\end{equation}
recalling Eqs.\ (\ref{eq_gov},\ref{galerkin}) and performing the integrals over an element, we obtain

\begin{equation}
	\mbf{M}_{(\trm{e})} \ddot{\bsy{q}}_{(\trm{e})}+\mbf{K}_{(\trm{e})}\bsy{q}_{(\trm{e})}+\bsy{Q}_{(\trm{e})}=\bsy{0},
\end{equation}
where ${M_{(\trm{e})}}_{ij}=\int_{0}^{h_{\trm{e}}}\rho A {\psi_{(\trm{e})}}_i(x) {\psi_{(\trm{e})}}_j(x) \dd x $,  ${K_{(\trm{e})}}_{ij}= \int_{0}^{h_{\trm{e}}}E I {\psi_{(\trm{e})}^{''}}_{i}(x) {\psi_{(\trm{e})}^{''}}_{j} (x)  \dd x $. The generalised force corresponding to the coordinate ${q_{(\trm{e})}}_i(t)$ is given by 
$$ {Q_{(\trm{e})}}_i=\gamma_{\trm{h}}\int_{0}^{h_{\trm{e}}} z_{(\trm{e})}(x,t) {\psi_{(\trm{e})}^{''}}_{i}(x) \dd x .$$
Letting $x=\frac{h_{\rm e}}{2}\left(1+\zeta\right)$,
\begin{align}
	{Q_{(\trm{e})}}_i&=\frac{\gamma_{\trm{h}} h_{\trm{e}}}{2} \int_{-1}^{1} {z}_{(\trm{e})} \left(h_{\trm{e}}\frac{1+\zeta}{2},t\right)  {{\psi}_{(\trm{e})}^{''}}_{i} \left(h_{\trm{e}}\frac{1+\zeta}{2}\right) \dd \zeta\, \nonumber\\
	&=\frac{\gamma_{\trm{h}} h_{\trm{e}}}{2} \int_{-1}^{1} \overline{z}_{(\trm{e})}(\zeta,t){\overline{\psi}_{(\trm{e})}^{''}}_{i}(\zeta) \dd \zeta \nonumber \\
	&=\frac{\gamma_{\trm{h}} h_{\trm{e}}}{2}  \sum_{p=1}^{n_{\trm{g}}} w_p \;\overline{z}_{(\trm{e})} (\zeta_p,t) \; {\overline{\psi}_{(\trm{e})}^{''}}_{i} \left(\zeta_p\right)\nonumber  \,\trm{(Gauss quadrature)}\\
	&=\frac{\gamma_{\trm{h}} h_{\trm{e}}}{2}  \sum_{p=1}^{n_{\trm{g}}} {\overline{\psi}_{(\trm{e})}^{''}}_{i} \left(\zeta_p\right) \;{\overline{z}_{(\trm{e})}}_p\;  w_p
\end{align}
where  $ {\overline{z}_{(\trm{e})}}(\zeta,t)=z_{(\trm{e})}\left(h_{\trm{e}}\frac{1+\zeta}{2},t\right) $, ${\overline{z}_{(\trm{e})}}_p={\overline{z}_{(\trm{e})}}(\zeta_p,t),$ ${\overline{\psi}_{(\trm{e})}^{''}}_{i}(\zeta)={\psi_{(\trm{e})}^{''}}_{i}\left(h_{\trm{e}}\frac{1+\zeta}{2}\right)$. The $\zeta_p\mbox{ and }w_p \:(p=1,2,\dots n_{\trm{g}})$ are the Gauss integration points and their corresponding weights.

The dynamics of hysteretic dissipating moments at the Gauss points within an element are governed by the equations

\begin{multline}
	\ddx{t}{{\overline{z}}_{(\trm{e})}}_p= \left(\bar{A}-\alpha\: \trm{sign}\left({{\overline{{z}}}_{(\trm{e})}}_p \ddx{t}{\chi_{(\trm{e})}}_p\right)\abs{{{\overline{{z}}}_{(\trm{e})}}_p}^{n_{\trm{h}}}  -\beta \abs{{{\overline{{z}}}_{(\trm{e})}}_p}^{n_{\trm{h}}}  \right) \times\\
	\ddx{t}{{{\chi}}_{(\trm{e})}}_p,\;\;p\in\{1,2,\dots,n_{\trm{g}}\}
\end{multline}
where ${\chi_{(\trm{e})}}_p=\sum_{i=1}^{4}{\overline{\psi}_{(\trm{e})}^{''}}_{i}(\zeta_p) {q_{(\trm{e})}}_i(t) $ is the curvature at the $p^{\trm{th}}$ Gauss point.
\begin{equation}
	\bsy{Q}_{(\trm{e})} =\mbf{A}_{(\trm{e})}\overline{\bsy{z}}_{(\trm{e})}, \,\bsy{\chi}_{(\trm{e})}={\overline{\bsy{\Psi}}_{(\trm{e})}}^{\top}\bsy{q}_{(\trm{e})},\, \mbf{A}_{(\trm{e})}=\frac{\gamma_{\trm{h}}\, h_{\trm{e}}}{2} \overline{\bsy{\Psi}}_{(\trm{e})}\;\mbf{W},
\end{equation}
where 
\begin{align}\label{mat}
	\left[\overline{\bsy{\Psi}}_{(\trm{e})}\right]_{ip}&={\overline{\psi}_{({\rm e})}^{''}}_{i}(\zeta_p),\, {\bf W}={\rm diag}([\zeta_1,\zeta_2,\dots,\zeta_{n_{\rm g}}]), \nonumber \\
	\overline{\bsy{z}}_{({\rm e})}&=[{\overline{z}_{(\trm{e})}}_{1},{\overline{z}_{(\trm{e})}}_{2},\dots,{\overline{z}_{(\trm{e})}}_{n_{\rm g}}]^{\top},\,
	i\in\{1,2,3,4\},\mbox{ and }\nonumber\\
	p&\in\{1,2,\dots,n_{\rm g}\}.
\end{align}

\small{
}

\end{document}